\documentclass[review]{elsarticle}

\journal{Journal of \LaTeX\ Templates}

\usepackage{xspace}
\usepackage{enumerate}
\usepackage{enumitem}
\usepackage{subfig}
\usepackage{multirow}
\usepackage{amsmath}
\usepackage{epstopdf}
\usepackage{appendix}
\usepackage{amsfonts}
\usepackage{color}
\usepackage[ruled]{algorithm2e}

\newtheorem{definition}{Definition}

\newtheorem{ccase}{Case}

\usepackage{threeparttable}
\usepackage{dcolumn}
\newcolumntype{d}[1]{D{.}{.}{#1}}% or D{.}{,}{#1} or D{.}{\cdot}{#1}

\newcommand{\eat}[1]{}
\newcommand{\paratitle}[1]{\vspace{1ex}\noindent \textbf{#1}}

\let\oldhat\hat
\renewcommand{\vec}[1]{\mathbf{#1}}
\renewcommand{\hat}[1]{\oldhat{\mathbf{#1}}}
\renewcommand{\matrix}[1]{\mathbf{#1}}

\newcommand{\eg}{\emph{e.g.,}\xspace}

\newcommand{\ie}{\emph{i.e.,}\xspace}

\newcommand{\etal}{\emph{et al.}\xspace}

\begin{document}
\begin{frontmatter}

\title{User-based Network Embedding for Collective Opinion Spammer Detection}
% \tnotetext[mytitlenote]{Fully documented templates are available in the elsarticle package on \href{http://www.ctan.org/tex-archive/macros/latex/contrib/elsarticle}{CTAN}.}

% Group authors per affiliation:
\author[HUST]{Ziyang Wang}
%\ead{ziyang1997@hust.edu.cn}

\author[HUST]{Wei Wei\corref{correspondingAuthor}}
\cortext[correspondingAuthor]{Corresponding author}
\ead{weiw@hust.edu.cn}

\author[BIT]{Xian-Ling Mao}
%\ead{maoxl@bit.edu.cn}

\author[NEU]{Guibing Guo}
%\ead{guogb@swc.neu.edu.cn}

\author[HUST]{Pan Zhou}
%\ead{panzhou@hust.edu.cn}

\author[UAE]{Shanshan Feng}
%\ead{victor_fengss@foxmail.com}

\address[HUST]{Cognitive Computing and Intelligent Information Processing (CCIIP) Laboratory,
School of Computer Science and Technology, Huazhong University of Science and Technology, China}
\address[BIT]{School of Computer Science and Technology, Beijing Institute of Technology, China}
\address[NEU]{Software College, Northeastern University, China}
\address[UAE]{Inception Institute of Artificial Intelligence, UAE}

\begin{abstract}
%Online reviews have become an important source of reference for decision-making. Positive reviews mean more users' favor while negative reviews may lead to business failure.
%
Due to the huge commercial interests
%the high profit it offers to fraudsters
%
behind online reviews, a tremendous amount of spammers manufacture
spam reviews for product reputation manipulation.
%(\ie promote/ defame).
% target products to promote or demote their reputations.
%
To further enhance the influence of spam reviews, spammers often collaboratively post spam reviewers within a short period of time,
%which is defined as \emph{collective opinion spam campaign}.
the activities of whom are called \emph{collective opinion spam campaign}.
%
%Recently, there exist numerous spam campaigns in platforms supporting online reviews.
%As the dynamically change in goals and members of such spam campaign activities
As the goals and members of the spam campaign activities change frequently,
and
%some spammers are also required to balance their workload within spam campaigns to evade detection,
some spammers also imitate normal purchases to conceal identity,
%and they often balance workload within spam campaigns for evading detection,
which makes the spammer detection challenging.
%the detection of spammers difficult.
%Hence, previous work using \emph{linguistic} cues or \emph{shallow behavioral} footprints are insufficient for capturing collusion signals. Intuitively, colluders are more likely to share more common neighbors within the same spam-campions (even without direct collusion relation), and thus the combination of \emph{direct} and \emph{indirect} neighbors of users is a relatively stationary collusion signal, which is not easily manual manipulation, unless no spammers are rehired.
% Thereby, i
In this paper, we propose an \emph{unsupervised} network embedding-based approach to
%jointly and seamlessly integrate
jointly exploiting different types of relations,
\eg direct common behaviour relation
%(called \emph{direct} relation)
and
indirect co-reviewed relation
%(refers to indirect associations with multiple-steps, \eg $k$-steps, along with the neighborhood structures of users, named \emph{indirect} relation),
to effectively represent the relevances of users for detecting the collective opinion spammers.
%we explored the relationship between users from multiple perspectives. For those who have direct common behaviour, a signed network is used to construct their direct connectivity. And for all the user in rating platforms, we use a global perspective to mine their potential relation. By jointly optimizing indirect and direct correlations, users can be effectively embedded in a low-dimensional space.
%
%{\color{red}
%Extensive experiments conducted on two real-world datasets show the significant improvements on different evaluation metrics,
The average improvements
of our method over the state-of-the-art solutions
on dataset \emph{AmazonCn} and \emph{YelpHotel} are [14.09\%,12.04\%]
and [16.25\%,12.78\%] in terms of \emph{AP} and \emph{AUC}, respectively.
%BLEU-4 / CIDEr / SPICE scores are 0:372, 1:226 and 0:216, respectively,
%as compared to the state-of-the-arts.
%}
%demonstrate the effectiveness of our proposed method significantly outperform the state-of-the-art baselines.
% To address this problem, in this paper we propose a graph-based network embedding based approach to estimating the probability of users participating in a spam campaign.
%
%Further, we explored the relationship between users from multiple perspectives. For those who have direct common behaviour, a signed network is used to construct their direct connectivity. And for all the user in rating platforms, we use a global perspective to mine their potential relation. By jointly optimizing indirect and direct correlations, users can be effectively embedded in a low-dimensional space.
%
%We demonstrate the effectiveness and stability of the method in two real-world datasets. The direct and indirect methods are also testified to be effective.}
\end{abstract}

\begin{keyword}
Spam Detection \sep Collective Spammer \sep Network Embedding \sep Signed Network
% \MSC[2010] 00-01\sep  99-00
\end{keyword}

\end{frontmatter}

% \linenumbers

\section{Introduction}
%Review spam detection (\aka \emph{Opinion Spam Detection}) is a subclass of information filtering techniques that trade-offs between incorrectly rejecting legitimate instances (\emph{false positives}) as opposed to not rejecting all spams (\emph{false negatives}), which actually aims to identify and remove malicious opinions (\ie human-powered deceptive contents), and thus significantly alleviate the negative impact of such reviews.

Online reviews are valuable source of reference for decision-making,
and thus massive volume of spammers are attracted to post malicious generated reviews to promote/demote the target products.
As opposed to individual spammers, collusive spammers are more harmful as they can easily propagate deceptive reviews for dominating the opinions of target products within a short period,
and their activities are called \emph{collective opinion spam campaign}.
As such, it is highly valuable and desirable to develop an effective algorithm  in accurately capturing collusion signals for collective opinion spammer detection.
The review spam detection problem has attracted extensive attentions.
Although these methods have greatly  advanced the problem, they still easily fail due to the facts:
\emph{Firstly}, previous \textbf{\emph{supervised}}-based methods formulate it as a classification task, whose performance often heavily rely on the scale of labeled data \cite{fornaciari2014identifying,ren2016deceptive,wang2016learning,li2017bimodal}.
Nevertheless, manual labeling spam reviews is extremely difficult \cite{ott2011finding}.
\emph{Secondly}, there exist several attempts for the spammer detection problem in an \textbf{\emph{unsupervised}} fashion:
(i) Previous \emph{linguistic}-based methods typically have poor performance as the reviews are usually manipulated by spammers
who would fine-tune their language~\cite{ye2015discovering,mukherjee2013yelp};
and (ii) Many \emph{behavior}-based approaches focus on modeling highly-visible behaviors~\cite{lim2010detecting,jindal2010finding,li2011learning,li2017bimodal} or temporal factors~(\eg co-bursting)~\cite{xie2012review,gunnemann2014detecting,gunnemann2014robust,li2015analyzing,kc2016temporal,ye2016temporal} for detecting spammers.
However, some spammers also make normal purchases to conceal identity, and thus such features are also not reliable for tackling the spammer detection problem.

Recently,  exist several efforts have been dedicated to research on spammer detection by modeling the user-group-product relations over behavior features~\cite{li2014towards,you2018attribute,kumar2018rev2}. These efforts usually focus on
extracting discriminative features (\ie pointwise or pairwise features) from \emph{direct} common behaviour relations (which are called direct relations).
However, spammers are often required to balance workload within spam campaigns for evading detection, and thus it is insufficient to capture the collusion signals by modeling the direct relation alone.
Intuitively,
spammers are usually engaged repeatedly for different collective opinion spam campaigns and colluders are likely to share more common neighbors within the same spam-campaigns (even without direct collusion relation), and thus the combination of the \emph{direct} and \emph{indirect} collusion relation (which refers to indirect associations with multiple steps, \eg $k$-steps, along with the neighborhood structures of colluders, which are called \emph{indirect} relation) is a relatively stationary collusion signal, which is not easily manually manipulated.

To this end, in this work we propose a \emph{unsupervised} network embedding-based approach to learning the user embeddings by jointly exploiting the different
types of relations (\ie \emph{direct} relation and \emph{indirect} relation) between pairwise users for \emph{collective opinion spam detection}.

The main contributions of our proposed method can be summarized as follows: 
i) \textbf{To be our best knowledge}, we are the first to jointly learning the direct relevance and indirect relevance for collective spammers detection;
ii) We develop a signed network composed of positive and negative links,
in which the direct relevance measures the degree of being colluders in the signed network,
and the indirect relevance measures the potential possibility of being colluders via a k-step neighborhood proximity over the signed network;
iii) Extensive experiments conducted on two real-world datasets demonstrate our proposed method significantly outperforms the state-of-the-arts. 
% Specifically,
% the \emph{direct} relation is used to learn \emph{direct relevance} embeddings for predicting
% the degree of being colluders based on the direct co-rating relationships of any given pair of  users,
% while the \emph{indirect} relation is to learn \emph{indirect relevance} embeddings based on  a $k$-step co-rating neighborhood proximity
% between a pair of users.

\section{Related Work}
%
%The spam detection problem has been studied broadly in different domains, \eg
%extensive attentions of researchers in various domains, such as
%email.
%~\cite{cikm:chirita2005mailrank},
%
%Web,
%~\cite{sigir:castillo:2007:know,sigkdd_exp:spirin:2012:survey},
%
%blog,
%~\cite{aaai:kolari2006detecting},
%
%social networks
%~\cite{sigir:lee2010uncovering,aaai:zhu2012discovering,ijcai:hu2013social},
%
%community-driven question answering
%~\cite{www:liu2017detecting}
%
%and \etc
%
%Nevertheless, few of these studies touch the opinion spam detection problem, which dynamics differently.
%In this section, we mainly focus on reviewing 
We categorize the the existing work on
the opinion spam detection 
%problem,
%
%which relates with three research areas:
%
%which can be roughly categorized 
into two groups,
\ie \emph{linguistic}-based, \emph{behavior}-based.
%\emph{collective spam detection}, \emph{individual spam review detection} and \emph{network embedding}.
%
%Next, we will present an overview of the most related work in each area.
%which can be roughly categorized into three groups, \ie \emph{linguistic}-based, \emph{behavior}-based and \emph{network embedding}-based.

\paratitle{Linguistic-based Spam Detection}.
Linguistic-based methods aim at extracting the discriminative linguistic features to differentiate the fake users
from normal ones.
%from linguistic perspective.
%
For example, these methods 
%Ott \etal\shortcite{ott2011finding} 
identify review spams according to
linguistic clue~\cite{ott2011finding, shehnepoor2017netspam, liu2019opinion},
writing-style feature~\cite{harris2012detecting},
syntactic pattern~\cite{feng2012syntactic},
LDA-based topic model~\cite{li2014spotting},
Bayesian generative model~\cite{li2014towards},
positive-unlabeled learning~\cite{li2014spotting},
frame-based model~\cite{kim2015deep},
and document-level features~\cite{ren2016deceptive}.
%
%on users' comments from Psychological aspect.
% and they also contribute the first ground truth dataset based on a crowd-sourcing platform named  \emph{Amazon Mechanical Tuck}\footnote{\url{https://www.mturk.com/}.}.
%
%Harris~\shortcite{harris2012detecting} explore writing-style features for spotting deceptive opinion spam
%several assessment methods based on writing style features to spot deceptive opinion spam.
%
%Feng~\etal~\shortcite{feng2012syntactic} investigate  syntactic patterns for review spam detection.
%syntactic stylometry
%
%
%Li~\etal propose a LDA-based topic model~\cite{li2013topicspam} and a Bayesian generative approach~\cite{li2014towards} for discovering the difference between deceptive reviews and truthful ones.
%Later on, they further propose a Bayesian generative approach to discovering  a profound discrepancy between deceptive reviews and truthful ones~\cite{li2014towards}.
%
%Li~\etal~\shortcite{li2014spotting} propose a positive-unlabeled learning approach to explore uni-grams and bi-grams features.
% for spam detection.
%
%Kim \etal \shortcite{kim2015deep}  propose a frame-based approach to analyze the characteristics of deceptive  in semantic-level for deceptive opinion detection.
%and truthful opinions
%
%A gated recurrent neural network proposed by \cite{ren2016deceptive} to model document-level features for review spam identification.
%representation to directly form it as features for deceptive opinion spam identification.
%
However, Mukherjee \etal \cite{mukherjee2012spotting} show that linguistic-based features are ineffective for spam detection problem.
In contrast, our proposed model 
%is mainly 
focuses on modeling spam-campaign characteristics (\textbf{\emph{excluding}} textual features) within a unified \textbf{\emph{unsupervised}} framework to explore the direct/indirect \textbf{\emph{user-user}} relations of users
to capture the collusion signals.
% for detecting collective opinion spammers.
%
%to simultaneously leverage \textbf{\emph{direct}} connected users (called \emph{direct} relation) and \textbf{\emph{indirect}} relations~(\ie indirectly connected users, called \emph{indirect} relation) to capture the collusion signals for detecting collective opinion spammers.

\paratitle{Behavior-based Spam Detection}.
Behavior-based spam detection aims at detecting a set of \emph{collective} malicious manipulation of online reviews
%by capturing the collusion signals
according to behavior-based features.
Lim \etal~\cite{lim2010detecting} study the rating behavioral characteristics for review spammer detection.
Jindal \etal~\cite{jindal2010finding} find unexpected rules to represent suspicious behaviors of reviews based on
%the identification of
the \emph{unusual} review patterns.
Li~\etal~\cite{li2011learning} propose a co-training method based on the extracted review-based/reviewer-based features for identifying spammers.
%
%Xie \etal (2012) explore the singleton reviews with abnormal temporal patterns.
%
Li \etal \cite{li2017bimodal} propose to employ a coupled hidden Markov model to discover co-bursting behavior patterns.
% among users for spammer detection.

Li \etal \cite{li2019unsupervised} focus on the cold-start problem by using user behavior representation.
Several efforts~\cite{xie2012review,gunnemann2014detecting,gunnemann2014robust,li2015analyzing,kc2016temporal,ye2016temporal} have also dedicated to research on temporal factors
% to caputure bursty signals
for spam detection.
%
%such as abnormal temporal patterns~\cite{xie2012review}, temporal dynamics~\cite{}, rating time-series~\cite{,}
However, some spammers also make normal purchases to conceal identity, which
affects the effectiveness of these 
%
%significantly affect the importance of 
spammer detection methods.
%makes the problem more challenging.
Hence, the approaches 
that heavily reply on \emph{shallow behavior features} or \emph{temporal factors}  are insufficient for accurately identifying the spammers.

There have been numerous attempts to detect spammers by modeling the user-group-product relations over behavior features.
Mukherjee \etal \cite{mukherjee2012spotting}
%propose a relation-based model to
rank collusion spam groups over user-group-product relations for detection.
Ye \etal \cite{ye2015discovering} estimate the likelihood of products being spam campaign target for inferring spammers.
% based on the estimation.
%clustering spammers induced by the most likely target products.
%, which first estimate the likelihood of products being spam campaign targets and then cluster spammers induced by the most likely target products.
%
Rayana \etal \cite{rayana2015collective} utilize a loopy belief propagation model to infer spammers by extracting relational features, which is then extended by introducing active inference\cite{rayana2016collective}.
Wang \etal \cite{wang2016learning} learn the user embeddings generated by tensor decomposition for training a
spam review classifier.
Liu \etal \cite{liu2017holoscope} mainly focus on graph topology and temporal information for detecting fraud users.
Xu \etal \cite{xu2017online} propose a regularized matrix factorization model to obtain reviewer behavior embeddings.
Kaghazgaran \etal \cite{kaghazgaran2018combating} pre-train a spammer classifier by modeling structurally similar users within a three-phase framework.
Kumar~\etal~\cite{kumar2018rev2} iteratively calculate three quality metrics
for spammer detection.
Wang~\etal~\cite{Wang2020} design a loopy belief propagation based algorithm to detect spammer groups.
% based on the intuition of normal users being inclined to provide high quality of comments that are consist with the the goodness of products.
%
Additionally, You \etal \cite{you2018attribute} develop a unified deep learning architecture to tackle the cold-start problem in spam review detection.
% by enriching the existing entities with other domain knowledge.
%
However, the spam campaigns might balance workloads of spammers to evade detection.
% which makes problem more challenging.
%
%some spammers are often required to balance workload within spam campaigns for evading detection
%some spammers also make normal purchases to conceal identity, which makes problem more challenging.
%Thereby, previous work 
These existing methods use \emph{shallow relational} information, \ie directly connected users (called \emph{direct} relation) while neglecting \textbf{\emph{indirect}} relations~(\ie indirectly connected users, called \emph{indirect} relation)
, and they easily fail for 
%to effectively achieve 
the spammer detection task. %and there is no previous work to combine the \emph{direct} and \emph{indirect} \emph{user-user} relations.
%
% modeling the directly connected users (called \emph{direct} relation) while neglecting \textbf{\emph{indirect}} relations~(\ie indirectly connected users, called \emph{indirect} relation), and there is no previous work to simultaneously leverage such two relations for spammer detection.
%
%Intuitively, colluders more likely share more common neighbors that have participated within a same spam-campions (even they don't have any direct correlations), and thus the combination of direct/indirect neighbors is a relatively stationary collusion signal for collective review spammer detection, which \emph{is not easily manual manipulation, unless no spammers are rehired}. Hence, 
In contrast, we focus on modeling user behavioral information within a unified \textbf{\emph{unsupervised}} architecture to explore the direct/indirect \textbf{\emph{user-user}} relations among reviewers
%to capture the collusion signals
for detection.

\section{Problem Statement and Notations}
\label{sec-pro-def}

Let $\matrix{P}=\{p_j\}_{|\matrix{P}|}$ be the set of items over a set of product categories $\matrix{C}=\{c_i\}_{|\matrix{C}|}$;
$\matrix{U}=\{u_i\}_{|\matrix{U}|}$ and $\matrix{X}=\{\vec{x}_{ij}\}_{|\matrix{X}|}$ be the set of users
and the records of their reviews, where $\vec{x}_{ij}\in \matrix{X}$ denotes the reviews posted by $u_i$ on product $p_j$,
since a user $u_i$ might post different reviews for the same product $p_j$ owing to multiple purchases.
As such, a $4$-tuple $(u_i,p_j,r,t)$ denotes the review generated by user $u_i$
for product $p_j$ with the rating $r$ at time $t$, in which the rating and the time-stamp
are denoted by $x^{r}_{ijk}$ and $x^{t}_{ijk}$ for simplicity.

Then, given a set of users and the metadata of their posted reviews, \ie $(u_i,p_j,r,t)$, the problem of \emph{collective opinion spammer detection}
is defined as identifying a set of spammers based on their participated \textbf{\emph{spam campaigns}}, namely, which is regarded as a task that requires
a set of spammers to collectively post malicious opinions (\ie human-powered deceptive contents) on the target items,
and outputs a ranking list of candidates with the spamicity scores, which can be regarded as the likelihood of candidates participating in opinion spam campaign.

%\begin{definition}
%\label{definition-1}
%\textbf{Spam Campaign}.
%\emph{A spam campaign is a task that asks a set of spam reviewers to collectively post malicious opinions (\ie human-powered deceptive contents) on the target items.
%%within a short time frame.
%}
%\end{definition}
%
%Based on the definition, given a set of users and their posted reviews, the problem of \emph{opinion spammer detection} is to identify the spam reviewers
%from normal ones based on their participated \textbf{\emph{spam campaigns}}.

\begin{figure*}
  \centering
  \includegraphics[width=\columnwidth, angle=0]{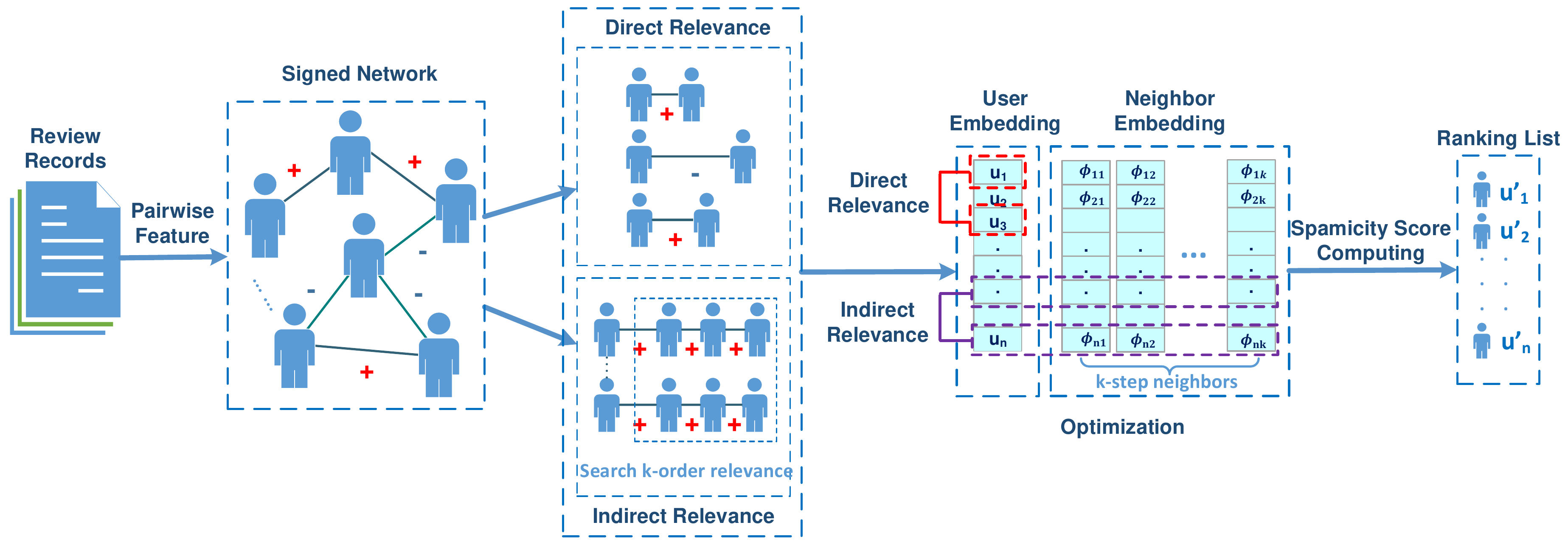}
 \vspace{0cm} \caption{\textbf{Overview of proposed approach}. First, a signed network is built based on the extraction of pairwise features from users' reviews, then we jointly optimize two types of relevances, \ie \emph{direct} relevance and \emph{indirect} relevance for learning the user embedding in a low-dimensional space.
  Finally, the spamicity score of each user is calculated based on the learnt user embeddings for estimating the degree of being colluders.}
  \label{fig:overview}
  \vspace{0cm}
\end{figure*}

\section{The Proposed Approach}

In this section, we present our proposed collective opinion spammer detection
approach within an unified architecture (shown in Figure \ref{fig:overview}), named \textsf{COSD}, the objective of which is to
jointly combine \emph{direct} and \emph{indirect} neighborhood exploration for learning the embedding representation of each user
for more accurately identifying spam reviewers.
%of neighborhood characteristics for learning the mixture embeddings of each user to more accurately identify spam reviewers.
The rationale behind is that
\emph{direct relevance embedding} controls the learning of user embeddings towards the pairwise users with a strong intensity of the collusive characteristics, while \emph{indirect relevance embedding} tends to make pairwise users sharing more commonly co-rating neighbors closer.
%during learning.
%
Indeed, we are only interested in \emph{direct relevance} embedding of users to identify spammers, however we also consider
\emph{indirect relevance} embedding in our framework because such two types of embeddings will reinforce each other mutually
to make the relevant users close over \emph{direct} and \emph{indirect} relations of users (the details will be elaborated later on).
%
%More specifically, \emph{direct embedding} tends to learn embeddings for predicting the degree of being colluders based on the direct co-rating relationships of any given pairwise users, while the \emph{indirect  embedding} is to learn embeddings based on  a $k$-order co-rating neighborhood proximity between any pairwise users. Both of such two types of relevance embeddings are simultaneously learned and  modeled into an unified objective function for learning user embeddings in a theoretically sound way.
%
Next, we detail how to model the \emph{direct relevance embedding} and \emph{indirect relevance embedding} for any pairwise users.

\subsection{Modeling Spam-Campaign Characteristics}
\label{sec-fc-context}
%
%Many previous studies are proposed to characterize spam-campaigns on different dimensions~\cite{mukherjee2012spotting,xu2015combating,xu2017online,wang2016learning}.
%
In this section, we mainly focus on how to characterize spam-campaigns on different dimensions for building the user-user weighted matrix $\matrix{W}$
to capture the intersections on the co-rated item sequences for each pairwise users.
Many previous works are proposed for this task~\cite{mukherjee2012spotting,xu2015combating,xu2017online,wang2016learning}.
By following the work~\cite{xu2015combating}, we adopt four different types of heterogeneous
pairwise features as follows.
Note we do not consider \emph{linguistic}-based features from reviews.
%~\cite{mukherjee2012spotting}.
%, as it is not effective for reflecting traits of spammers~\cite{mukherjee2012spotting,ye2015discovering}.
% may fail to reflect traits of spammers who would  fine-tune their language~\cite{ye2015discovering}.
%
\begin{itemize}
    \item \textbf{Product Rating Proximity~(\textsf{PR})}, it measures the intensity of spammers' agreements for pairwise users.
    % over their co-rating relations.
    Generally, spammers within a spam-campaign are instructed to post similar
    %(\eg promote/demote)
    opinions with consistent ratings on target items.
    %~\cite{xu2015combating}.
    As such, given a pair of users ($u_i,u_j$), we
    measure the intensity of spammers' agreements
    as follows,
    \begin{equation}
    \label{eq-psi-pr}
    \psi_{PR}(i,j)  = \frac{2}{1+\exp(\Gamma^{p,r}_{ij})}
    \end{equation}
    where $\Gamma^{p,r}_{ij}$ denotes the average rating deviation of pairwise users ($u_i,u_j$)
    over their commonly reviewed items,
    \begin{math}
      \Gamma^{p,r}_{ij}= \frac{1}{|\vec{P}_{i}\cap \vec{P}_{j}|}\sum_{p_k\in \vec{P}_{i}\cap \vec{P}_{j}}\left(\left|\frac{\sum_{q}x^{r}_{ikq}}{|x_{ik}|}-\frac{\sum_{q}x^{r}_{jkq}}{|x_{jk}|}\right|\right),
      %\overline{x^{r}_{ik}}-\overline{x^{r}_{jk}}
    \end{math}
    %
    %and $Z(i,j)$ is a normalizing factor formulated as
    where $\vec{P}_{i(j)}$ refers to the set of items reviewed by user $u_{i(j)}$.
    Eq. (\ref{eq-psi-pr}) favors to find pairwise spammers who have
    more consistent ratings on co-rated items,
    especially $\psi_{PR}(i,j)=1$ when $\Gamma^{p,r}_{ij}=0$.
    \item \textbf{Product Time Proximity~(\textsf{PT})}, it captures the temporal consistency for pairwise users. Intuitively, colluders are often asked to complete the task within a short-time frame (\eg less than a week) for maximizing the influence,
    % effectiveness of such spam campaign,
    and thus the temporal traces of their reviews tend to be more intensive than normal users'.
    Hence, we use \textsf{PT} to capture the temporal consistency of pairwise users ($u_i,u_j$),
    \begin{equation}
    \psi_{PT}(i,j) = \frac{1}{C+\gamma \Gamma^{p,t}_{ij}}
    \end{equation}
    where $\Gamma^{p,t}_{ij}$ denotes the average time deviation of pairwise user ($u_i,u_j$)
    over their co-reviewed items,
    \begin{math}
      \Gamma^{p,t}_{ij}= \frac{1}{|\vec{P}_{i}\cap \vec{P}_{j}|}\sum_{p_k\in \vec{P}_{i}\cap \vec{P}_{j}}\left(\left|\frac{\sum_{q}x^{t}_{ikq}}{|x_{ik}|}-\frac{\sum_{q}x^{t}_{jkq}}{|x_{jk}|}\right|\right);
      %\overline{x^{r}_{ik}}-\overline{x^{r}_{jk}}
    \end{math}
    $C$ and $\gamma$ denote the smoothing factor and the trade-off parameter, which are empirically set at $1$ and $20$, respectively.
    \item \textbf{Category Rating Proximity~(\textsf{CR})}, it measures the average category rating deviation between pairwise users. Intuitively, spammers from different spam-campaigns might have different rating distributions over reviewed categories, and thus the higher intersections of category rating distributions of two users are consistent, the more likely these two users are colluders.
    Hence, \textsf{CR} is computed based on
    the average category rating deviation $\Gamma^{c,r}_{ij}$ between pairwise users,
    \begin{equation}
    \psi_{CR}(i,j) = \frac{2}{1+\exp(\Gamma^{c,r}_{ij})}
    \end{equation}
    where
    \begin{math}
      \Gamma^{c,r}_{ij}=
      \frac{1}{|\vec{C}_{i}\cap \vec{C}_{j}|}
      \sum_{c_k\in \vec{C}_{i}\cap \vec{C}_{j}}
      {\left(\left|\overline{c^{r}_{ik}}-\overline{c^{r}_{jk}}\right|\right)},
    \end{math}
    and $\vec{C}_{i(j)}$ is the set of categories reviewed by user $u_{i(j)}$;
    $\overline{c^{r}_{ik}}$ denotes the average rating of user $u_{i}$ in the $k$-th category, which is calculated by
    \begin{math}
      \overline{c^{r}_{ik}}=\frac{1}{|c_{k}|}
      {\sum_{p_{j}\in c_{k}}{\frac{1}{|\vec{x}_{ij}|}\sum_{q}{x^{r}_{ijq}}}}.
    \end{math}
    \item \textbf{Category Time Proximity~(\textsf{CT})}, it measures the consistency of time distributions between pairwise users over co-reviewed categories.
    Analogous to \textsf{CR}, it is also a strong indicator to measure the degree of the collusive characteristics of pairwise users,
    which is estimated by,
    \begin{equation}
    \psi_{CT}(i,j) = \frac{1}{C+\gamma \Gamma^{c,t}_{ij}}
    \end{equation}
    where $C$ and $\gamma$ are empirically set at $1$ and $20$, respectively;
    \begin{math}
      \Gamma^{c,t}_{ij}=
      \frac{1}{|\vec{C}_{i}\cap \vec{C}_{j}|}
      \sum_{c_k\in \vec{C}_{i}\cap \vec{C}_{j}}
      {\left(\left|\overline{c^{t}_{ik}}-\overline{c^{t}_{jk}}\right|\right)},
    \end{math}
    and
    {\small
    \begin{math}
      \overline{c^{t}_{ik}}=\frac{1}{|c_{k}|}
      {\sum_{p_{j}\in c_{k}}{\frac{1}{|\vec{x}_{ij}|}\sum_{q}{x^{t}_{ijq}}}}.
    \end{math}
    }
\end{itemize}

\paratitle{Pairwise Feature Combination}.
%As discussed in \cite{xu2015combating,xu2017online}, each feature alone is insufficient to statistically significant to indicate manipulation of campaigns.
%
To estimate the intensity of spammer agreement for any pairwise users,
we follow the work~\cite{xu2015combating} to employ a convex combination of the mentioned pairwise proximities with a weighting vector $\vec{\alpha}$
for calculating $ \hbar_{ij}$,
%to estimate the intensity of spammer agreement for any pairwise users,
%
%and thus all such pairwise features with a weighting vector $\vec{\alpha}$ for a pair of users ($u_i,u_j$)
%
%Hence, a final combination of such pairwise proximities is needed for yielding a strong indicator to estimate the intensity of spammer agreement for any pairwise users.
%
%To this end, we employ a convex combination of all such pairwise features with a weighting vector $\vec{\alpha}$ for a pair of users ($u_i,u_j$), namely,
%
{\vspace{0pt}
\begin{equation}
\label{eq:alpha}
  \hbar_{ij}=\sum_{k}\alpha_{k}\psi_{(.)}(i,j),
\end{equation}
}
where each feature $\psi_{(.)}(i,j)$ (\ie \textsf{PR},\textsf{PT},\textsf{CR},\textsf{CT}) is normalized within $[0,1]$,
and $\sum_{k}\alpha_k=1~(\alpha_k\geq 0)$, k is the index for each pairwise feature.

\begin{figure}[!t]
  \centering
  \includegraphics[width=0.8\columnwidth, angle=0]
  {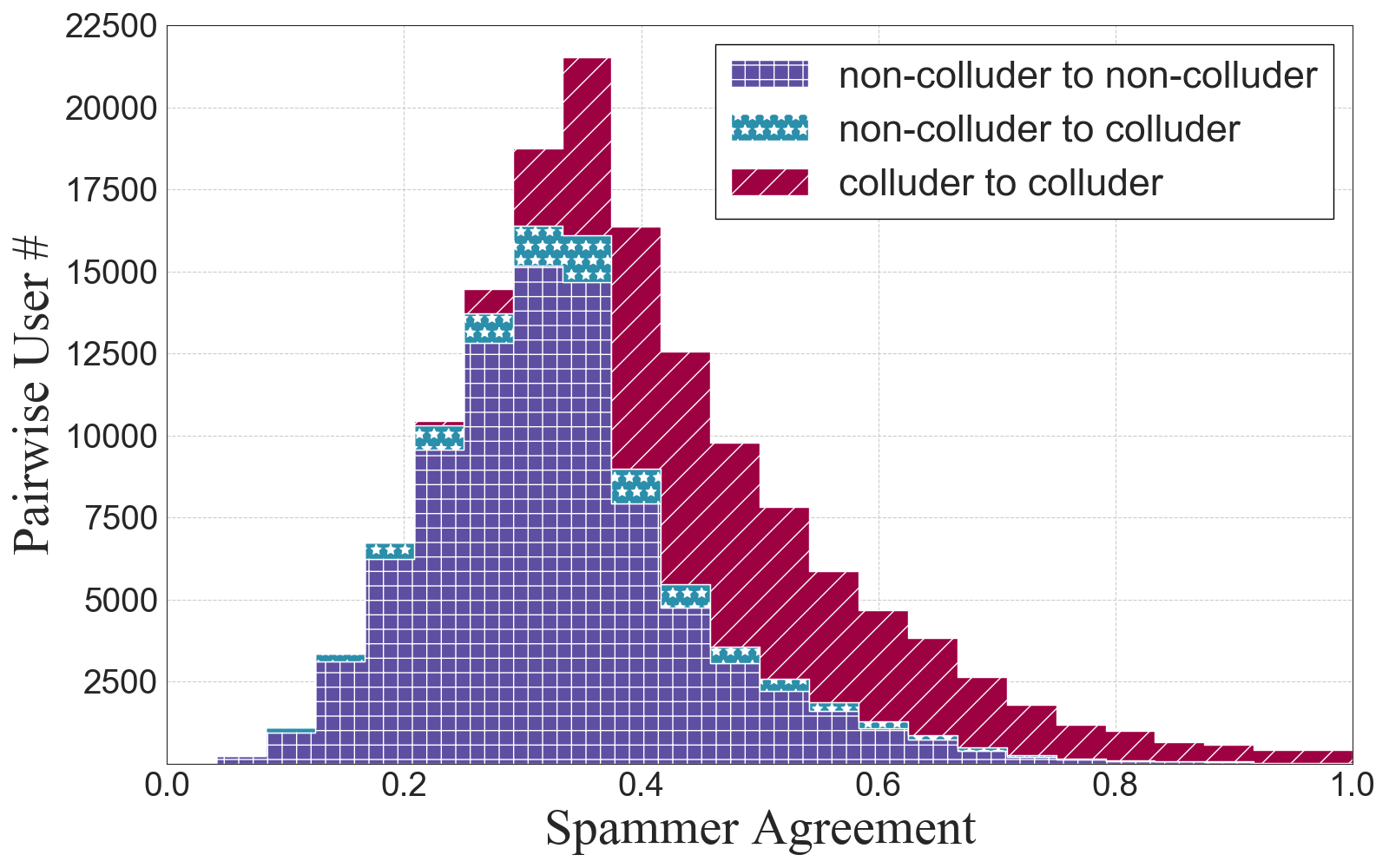}
  \vspace{0cm}
  \caption{\textbf{Empirically data analysis}.  We mainly focus on three different types of pairwise user relations collected from
  %  the change ratio of pairwise users in
  \emph{AmonzonCn}, \ie \textbf{C-C},
  %colluder to colluder;
  \textbf{NC-C},
  % non-colluder to colluder;
  \textbf{NC-NC}.
  %: non-colluder to non-colluder.
  %
  X-axis denotes the intensity of spammers' agreements for measuring the collusive characteristics between each pairwise users,
  and Y-axis indicates the corresponeding number of pairwise users.}
  \label{fig:emp_feature}
  \vspace{0cm}
\end{figure}

To verify the effectiveness of $\hbar_{ij}$,
we present an empirical data analysis on \emph{Amazon\_cn}~\cite{xu2013uncovering},
%from which we randomly sample $2,200$ users with $833$ colluders and $1,367$ non-colluders
%(\rf Section \ref{sec-dataset} for more details).
%
where we mainly focus on three different types of pairwise user relations according to their co-reviewed relations,
%build $3$ categories over pairwise users according to their co-reviewed items,
\ie \emph{colluder to colluder}~(\textbf{C-C}),
\emph{non-colluder to colluder}~(\textbf{NC-C}), and\emph{ non-colluder to non-colluder}~(\textbf{NC-NC}).
Figure \ref{fig:emp_feature} shows the distribution of the spammer-agreement ($\hbar_{ij}$) to the number of pairwise users
over such $3$ relations.
For the sake of discussion, we mainly focus on \textbf{C-C} and \textbf{NC-NC}.
As the \emph{spammer agreement} increases, the change ratio of \textbf{C-C} and \textbf{NC-NC} is different,
\ie the number of \textbf{NC-NC} considerably decrease occurred
within the range of $[0.4,0.8]$.
%from $\hbar_{ij}=0.4$ to $\hbar_{ij}=0.8$,
As opposed to \textbf{NC-NC}, \textbf{C-C} gradually decrease within such range.
In particular, the number of \textbf{NC-NC} almost equals to $0$ when $\hbar_{ij} \geq 0.8$,
and the number of \textbf{C-C} almost equals to $0$ when $\hbar_{ij}\leq 0.22$.
Similar trends are also reported in \cite{xu2015combating}.
Based on the observations, we
measure the collusive characteristics of each pairwise users as follows,
%update the measure of the collusive characteristics of each pairwise users as follows,
which is different from~\cite{xu2015combating},
%that is
%
\begin{equation}
w_{ij} =(\hbar_{ij} - \zeta) * \eta_{PI}(i,j),
\label{eq:weight}
\end{equation}
where $\zeta$ is a hyper-parameter that can be set in different ways, \eg $\zeta = \overline{\hbar_{(.)}}$, as most of pairwise users
belong to \textbf{NC-NC} and thus tend to be firstly filtered by our method;
$\eta_{PI}(i,j) = \frac{|P_{i}\cap P_{j}|}{\sqrt{|Pi|}\sqrt{|P_{j}|}}$ is a confidence score induced by the proportion
of items commonly reviewed by
%pairwise users
($u_i,u_j$).

%%%%%%%%%%%%%modification%%%%%%%%%%%%%%%%%%%%%%%%%%%%
%\paratitle{Pairwise Item Proximity~(\textsf{PI})}.
%%
%Normal users behave differently based on their individual preferences,
%thus intuitively posting more reviews by a pair of users on the common items
%is a strong indicator for measuring the collusiveness between them.
%%
%Given a pair of users $(u_i,u_j)$, \textsf{PI} is computed by the proportion
%of the commonly reviewed items, namely,
%%
%\begin{equation}
%\eta_{PI}(i,j)= \frac{|\vec{P}_{i}\cap \vec{P}_{j}|}{\sqrt{|\vec{P}_{i}|}\sqrt{|\vec{P}_{j}|}}
%\end{equation}
%%where $\vec{P}_{i(j)}$ refers to the set of items reviewed by user $u_{i(j)}$.
%It is worth noting we employ $\sqrt{|\vec{P}_{i}|}\sqrt{|\vec{P}_{j}|}$
%in estimating \textsf{PI}, rather than $\vec{P}_{i}\cup \vec{P}_{j}$ as \cite{xu2015combating},
%since the latter easily favors to the pairwise users purchasing more items
%with a higher value.

\subsection{Modeling of Direct Relevance Embedding}
\label{sec-direct}
In this section, we mainly focus on how to learn the \emph{direct} embeddings of users over \emph{direct} co-rating relations for making a pair of users with a strong intensity of being colluders closer, otherwise the ones with a weak intensity of ones far away.
As such, we first give a definition related to the \emph{direct relevance} of a pair of users.

\begin{definition}
\label{definition-2}
\textbf{Direct Relevance},
\emph{
which is used to measure the degree of the spammer agreements based on their direct co-rating associations for any given pairwise users ($u_i,u_j$).
%For any given pairwise users ($u_i,u_j$), the direct relevance is used to measure the degree of the spammer agreements based on their direct co-rating associations.
}
\end{definition}

%Definition \ref{definition-2} is used to measure the degree of spammer agreements between any two pairwise users who have directly co-reviewed correlations.
Sequentially, we define a \emph{user-based signed network} built based on direct relevance of users for embedding.% for learning user embedding.

\begin{definition}
\label{definition-3}
\textbf{User-based Signed Network}~(\textbf{USN}).
\emph{
A user-based signed network~\cite{leskovec2010predicting} is defined by a $2$-tuple, \ie $\mathcal{G}=(\matrix{U},\matrix{E})$, which consists of a set of users
$\matrix{U}=\{u_i\}_{|\matrix{U}|}$, as well as a set of positive links $\matrix{E}^{+}$ and a set of negative links $\matrix{E}^{-}$,
and $\matrix{E}=\matrix{E}^{+}\cup \matrix{E}^{-}=\{e_{ij}\}_{|\matrix{E}|}$.
}
\end{definition}

In \textbf{USN}, both positive and negative links are represented into a weighted matrix $\matrix{W}\in \mathbb{R}^{|\matrix{U}|\times |\matrix{U}|}$,
where each element $w_{ij}\in \matrix{W}$ indicates the intensity of being colluders
for a pair of users,
%
%is computed by Eq. (\ref{eq:weight}),
especially $w_{ij}=0$ denotes the missing link between $u_i$ and $u_j$.
Then, the \textbf{\emph{direct relevance}} can be estimated by employing a likelihood function to minimize the \emph{negative} log-likelihood
of collusion possibility for any pairwise users $(u_i,u_j)$,
{
\vspace{0pt}
\small
\begin{equation}
\label{eq:loss_dir_rel}
\mathcal{L}_{d}=
\mathop{\min}
{
\left(
-{\underset{e_{ij}\in \matrix{E}}{\sum}\!\!\log f(\vec{u}_i,\vec{u}_j;w_{ij})}
\right)},
\end{equation}
}
where $\vec{u}_{i(j)}$ denotes the $d$-dimensional user embedding vector.
$f(.,.;.)$ is a likelihood function, and
%$w_{ij}$ is the weight of link $e_{ij}$, which indicates the degree of being colluders for ($u_i,u_j$).
%
%In fact,
many approaches can be used to model it, here we define the function based on the principle of
%Rectified Linear Units, \ie
\emph{ReLu}~\cite{hahnloser2001permitted,hahnloser2000digital}, and Eq. (\ref{eq:loss_dir_rel}) can be rewritten as,

{
\vspace{0pt}
\small
\begin{equation}
\label{eq:loss_dir_rel_rewrt}
%\vec{\theta}^{*} \!\!&\leftarrow
\mathcal{L}_{d}=
\mathop{\min}
{
\left(
{
\underset{e_{ij}\in \matrix{E}}{\sum}
\!\! \max\left(0, w_{ij}\| \vec{u}_{i}-\vec{u}_{j} \|^{2}_{F} + \delta\right)
}
\right)},
\end{equation}
}
where $\|.\|^{2}_{F}$ is the Frobenius norm of a vector,
$\delta$ indicates a smoothing parameter.
As mentioned, the learnt user embeddings are expected to make a pair of users with a strong intensity of being colluders closer,
otherwise the ones with a weak intensity of ones far away.
More specifically, we consider the following cases regarding $w_{ij}$.
%according to the analysis in Figure \ref{fig:emp_feature}.

\begin{ccase}
\label{ccase-1}
\emph{$w_{ij} > 0$, which indicates $e_{ij}\in \matrix{E}^{+}$ and the learnt user embeddings should be more closer, otherwise it should be penalized
with a larger loss.}
\end{ccase}

\begin{ccase}
\label{ccase-3}
\emph{$w_{ij} < 0$, which means $e_{ij}\in \matrix{E}^{-}$ and is expected to keep the pairwise users away from each other in the learning space,
otherwise it should be penalized,
%\ie the learned embeddings with a small distance should be penalized with a loss,
and a hyper-parameter is used to control the penalty,
\ie the value of which should be set as $0$
%namely, it would be reduced to $0$
when
%the distance  $\delta$, \ie
$\|\vec{u}_{i}-\vec{u}_{j}\|^{2}_{F}> \frac{-\delta}{w_{ij}}$.}
\end{ccase}

According to the discussion, Eq. (\ref{eq:loss_dir_rel_rewrt}) can be rewritten by combining the above two cases,
{
\vspace{0pt}
\small
\begin{equation}
\label{eq:loss_dir_rel_rewrt_final}
\mathcal{L}_{d}=
\mathop{\min}
{
\left({
\underset{e_{ij}\in \matrix{E}}{\sum} %sum_{}
 I_{ij}w_{ij}\|\vec{u}_{i}-\vec{u}_{j}\|_{F}^{2}
}
\right)},
\end{equation}
}
where $\matrix{I} \in \mathbb{R}^{|\matrix{U}| \times |\matrix{U}|}$ is an indicator matrix with $I_{ij} = 0$ if $w_{ij} < 0 \land \|u_{i}-u_{j}\|_{F}^{2} > \frac{-\delta}{w_{ij}}$ and 1 otherwise.

Then, the remaining problem is how to estimate the intensity of being colluders,
as well as model the indirect relevance embedding for each pairwise users.
We will detail them  in the following sections, respectively.

\subsection{Modeling of Indirect Relevance Embedding}
\label{sec-indirect}
As mentioned,  spammers within the spam campaigns might balance their  workload to evade detection,
and thus explicitly modeling the \emph{direct} relevance via \emph{shallow relational} information (\ie direct co-rating relation) alone
is insufficient for accurately capturing the collusion signals,
%the \emph{direct} relation might be not enough for accurately capturing the collusion signals,
which might make matrix $\matrix{W}$ sparse and result in a poor performance in spammer detection,
even worse when no direct correlations.
Intuitively, users sharing more common neighbors may be a strong indicator that they are colluders,
and thus it is a natural way to combine the direct relation and indirect relation,
which is a relatively stationary collusion signal for collective review spammer detection, and is also not
easily manual manipulation.
%, unless no spammers are rehired.
%Explicitly modeling the direct relevance via direct connections between a pair of users is not sufficient, as not all of pairwise users have commonly co-reviewed items, which may result in a poor accurate performance due to the sparsity of matrix $\matrix{W}$. Intuitively, the pairwise users sharing more commonly co-reviewed neighbors may be a strong indicator that they are colluders.
%
%Note that here we consider not only  the direct connections among users (\ie $1$-hop), but also the indirect associations among them, \eg $k$-hops.
Hence, here we mainly focus on how to measure the \emph{indirect relevance} for user embedding,
via modeling \emph{indirect} common neighbors for a pair of users.

\subsubsection{Indirect Relevance Proximity}
%
%Intuitively, users that are colluders more likely share more common neighbors that have participated in the same spam-campions, even they don't have any direct correlations.
%
To measure the \emph{indirect} relevance for \emph{pairwise} users, we employ a truncated \emph{random walk}~\cite{perozzi2014deepwalk} to model the \emph{indirect} neighborhood network structure of such users over
%the built
\emph{USN}.
Note we solely consider to walk along with the positive links (\ie $\matrix{E}^{+}$)
for modeling the collusive characteristics of each pairwise users,
and which is thus called  \textbf{\emph{positive-based randomwalk}}.
%
%randomly moves $k$-steps on the built \emph{USN} from a given point based on the \emph{indirect} neighborhood network structure of such users.
%
%Note that here we need to obtain $k$ neighbors to represent each user by modeling the collusive characteristics for each pairwise users, and thus here we consider the walk solely along with the positive link (\ie $\matrix{E}^{+}$), which is called \emph{Positive}-based Randomwalk.
%
Specifically, we obtain $r$ sequences with the maximum length of $k$ steps for user $u_i$,
and thus totally obtain $r\times |\matrix{U}|$ sequences ($\matrix{S}^{+}$),
where the \emph{indirect} neighbors of user $u_i$ is collected from sequences containing the neighborhood structure of $u_i$,
in which the interval between the points and $u_i$ is within the window size $\omega$.

%\paratitle{Positive-based Randomwalk}.
%%
%To model the neighborhood network structure of each user, here we employ the truncated \emph{random walk}~\cite{bar2008random} method to randomly moves $k$-steps on the user-based signed network from a given point, which has been proven the effectiveness in modeling the topology of a graph.
%%
%Note that the obtained k-order neighbors is used to model the collusive characteristics for each pairwise users,
%and here we thus only consider one moving over the positive link (\ie $\matrix{E}^{+}$) is a valid walk.
%%
%Specifically, for each user $u_i$, we obtain $r$ sequences with the maximum length of $k$ steps,
%and thus we totally obtain $r\times |\matrix{U}|$ sequences ($\matrix{S}^{+}$),
%where the $k$-order neighbors of user $u_i$ is
%collected from sequences containing the neighborhood structure of $u_i$,
%in which the interval between the points and $u_i$ is within $\omega$ (\ie window size).

Then, we learn the \emph{indirect relevance} embeddings
%to measure the indirect relevance
over their common $k$-step neighbors.
%the embeddings of users that share more $k$-order neighbors are thus closely located in the learned vector space
%
Hence, we adopt \emph{skip-gram}~\cite{mikolov2013efficient} to compute the loss function of \emph{indirect relevance}, which is defined  by the sum of the negative log probability of any pairwise users based on the learnt \emph{indirect relevance} embeddings within a window size $\omega$.

%adopt the \emph{skip-gram} model~\cite{mikolov2013efficient} which is a powerful language model for predicting the context of a given word.

%Note that there are two different roles for modeling indirect relevance: 1) user itself and 2) neighbors of other users, and thus the embeddings of users that share more $k$-order neighbors are thus closely located in the learned vector space. Here, the loss function we adopt the \emph{skip-gram} model~\cite{mikolov2013efficient} which is a powerful language model for predicting the context of a given word. Therefore, the loss function is defined by the sum of the negative log probability of any pairwise users, that is,
{\small
\vspace{0pt}
\begin{equation}
\label{con:high_loss}
\mathcal{L}_{id}=
\mathop{\min}
{
\left({
\sum_{i = 1}^{|\matrix{U}|}
\!\!\!\!\underset{s \in \matrix{S}^{+} \atop u_i,u_j\in s}{\sum}
% \sum_{s \in \matrix{S}^{+};u_i,u_j\in s}
\!\!\!\!\left({
\underset{i-\omega < j < i+\omega}{\sum}
%\sum_{i-\omega < j < i+\omega}
\!\!\!\!\!\!\!\!-\log \Pr(\vec{u}_{j}|\vec{u}_{i})
}\right)
}\right)},
\end{equation}
}
where $\Pr(\vec{u}_{j}|\vec{u}_{i})$ is the co-occurrence probability parameterized using the inner product kernel with softmax,
which is defined by
{
\vspace{0pt}
\small
\begin{equation}
\label{con:high}
\Pr(\vec{u}_{j}|\vec{u}_{i}) = \frac{\exp(\vec{u}_{i}^{T}\vec{\Phi}_{j})}{\sum_{k=1}^{|\matrix{U}|}\exp(\vec{u}_{i}^{T}\vec{\Phi}_{k})}
\end{equation}
}
where
$\vec{u}_{i}$ denotes the learnt \emph{direct relevance} embedding of $u_i$;
and $\vec{\Phi}_{j}$ is $u_{j}$'s
\emph{indirect relevance} embedding of $u_{j}$ when $u_{j}$ is the \emph{indirect} neighbor of $u_i$,
which is based on the principle of making the users sharing more $k$ neighbors close in the learnt vector space.
% corresponding embedding vector when $u_{j}$ is treated as the learnt $k$ neighbors of $u_i$.

However, Eq.~(\ref{con:high}) cannot be scalable due to the expensive computation overheads,
as we need to finish the updates of all users when computing $\Pr(\vec{u}_{j}|\vec{u}_{i})$.
To tackle the problem, we employ negative sampling for optimization~\cite{mikolov2013efficient},
we formulate the negative sampling function for pairwise user ($u_i,u_j$) when computing Eq.~(\ref{con:high}),
{
\vspace{0pt}
\small
\begin{equation}
\log \sigma (\vec{u}_{i}^{T} \vec{\Phi}_{j}) + \kappa \mathbb{E}_{u_{n} \sim P_{n}(u)}[\log \sigma (- \vec{u}_{i}^{T} \vec{\Phi}_{n})]
\label{con:negativeSample}
\end{equation}
}
where $\sigma(x) = \frac{1}{1+\exp(-x)}$ is the sigmoid function; $\kappa$ is the number of negative samples.
We empirically set $P_{n}(u_i) \propto d_{u_i}^{3/4}$ as~\cite{mikolov2013efficient}, and $d_{u_i}$ is the degree of user $u_i$.

%\paratitle{Positive-based Randomwalk}.
%%
%To model the neighborhood network structure of each user, here we employ the truncated \emph{random walk}~\cite{bar2008random} method to randomly moves $k$-steps on the user-based signed network from a given point, which has been proven the effectiveness in modeling the topology of a graph.
%%
%Note that the obtained k-order neighbors is used to model the collusive characteristics for each pairwise users,
%and here we thus only consider one moving over the positive link (\ie $\matrix{E}^{+}$) is a valid walk.
%%
%Specifically, for each user $u_i$, we obtain $r$ sequences with the maximum length of $k$ steps,
%and thus we totally obtain $r\times |\matrix{U}|$ sequences ($\matrix{S}^{+}$),
%where the $k$-order neighbors of user $u_i$ is
%collected from sequences containing the neighborhood structure of $u_i$,
%in which the interval between the points and $u_i$ is within $\omega$ (\ie window size).
%%
%%the points from the sequences containing $u_i$ and the interval between such co-appearing points and $u_i$ is within a windows size $\omega$.

\subsection{The Unified Model}
\label{sec-um}
In this section, we present an unified model to optimize our \emph{collective opinion spammer detection} problem
by minimizing a combination loss which consists of three different loss terms, \ie two losses based on the two learnt relevance embeddings,
\ie \emph{direct relevance} embeddings and \emph{indirect relevance} embeddings, as well as a regularization loss,
which is formalized as,
%
%More specifically, \emph{direct embedding} tends to learn embeddings for predicting the degree of being colluders based on the direct co-rating relations of any  pairwise users, while the \emph{indirect embedding} is to learn embeddings based on a $k$ co-rating neighborhood proximity between any pairwise users.
%
%according to
%Eq.~(\ref{eq-framework}),
%Eq.~(\ref{eq:loss_dir_rel_rewrt_final}) and Eq.~(\ref{con:high_loss}),which is defined by
%
{
% \hspace{10mm}
\vspace{0pt}
\begin{equation}
\begin{split}
\mathcal{L}_{mix} &=
(1 - \beta) \mathcal{L}_{id}+ \beta \mathcal{L}_{d} + \psi \mathcal{L}_{reg} \\
&=
-(1-\beta)
%%%%%%%%%%
%\sum_{i = 1}^{|V|} \sum_{c \in C^{+}} \sum_{i-w < j < i+w} \log\frac{\exp(\boldsymbol{u}_{i}^{T}\boldsymbol{\Phi}_{j})}{\sum_{k=1}^{|V|}\exp(\boldsymbol{u}_{i}^{T}\boldsymbol{\Phi}_{k})}
%
{
\left({
\sum_{i = 1}^{|\matrix{U}|}
\!\!\!\!\underset{s \in \matrix{S}^{+} \atop u_i,u_j\in s}{\sum}
% \sum_{s \in \matrix{S}^{+};u_i,u_j\in s}
\!\!\!\!\left({
\underset{i-\omega < j < i+\omega}{\sum}
%\sum_{i-\omega < j < i+\omega}
\!\!\!\!\!\!\!\!\log \!
%\Pr(\vec{u}_{j}|\vec{u}_{i})
\frac{\exp(\boldsymbol{u}_{i}^{T}\boldsymbol{\Phi}_{j})}{\sum_{k=1}^{|V|}\exp(\boldsymbol{u}_{i}^{T}\boldsymbol{\Phi}_{k})}
}\right)
}\right)}\\
&+\beta
%\sum_{(i, j) \in E} I_{ij}w_{ij}\|\boldsymbol{u}_{i}-\boldsymbol{u}_{j}\|_{F}^{2} \\
\left(
{\underset{e_{ij}\in \matrix{E}}{\sum} I_{ij}w_{ij}\|\vec{u}_{i}-\vec{u}_{j}\|_{F}^{2}}
\right)
+ \psi(\|\matrix{U}\|_{F}^{2}),
\label{con:loss}
\end{split}
\end{equation}
}
where
$\beta$ is a trade-off parameter for controlling the contributions of the learnt \emph{direct} embeddings and \emph{indirect} embeddings;
$\|\cdot\|_{F}^{2}$ is the Frobenius norm of a matrix;
%a regularization term and
and $\psi$ is a regularization parameter.
Note that, for each user $u$, the direct user embedding $\vec{u}$ and the indirect user embedding $\vec{\Phi}$
are simultaneously learned during training.

\paratitle{Spamicity Score}.

The calculation of the final spamicity score
relies on the assumption that collective opinion spammer is 
based on jointly participated in a spam campaign via large-scale manipulation,
and thus users within a same campaign group
are more likely to have similar behaviors,
To this end, we employ the Frobenius distance~(similar to \cite{xu2017online}) \ie
\begin{math}
  Score_{F}(u_i,u_j)=\exp(-\|\vec{u}_{i}-\vec{u}_{j} \|_{F}^{2})
\end{math}
%
%
%
%
%
%For the calculation of the final spamicity score,
%%of each pairwise user ($u_i,u_j$),
%we employ the Frobenius distance~(similar to \cite{xu2017online}) \ie
%\begin{math}
%  Score_{F}(u_i,u_j)=\exp(-\|\vec{u}_{i}-\vec{u}_{j} \|_{F}^{2})
%\end{math}
to measure the intensity of a pair of users ($u_i,u_j$) being colluders. The higher the $Score_{F}(u_i,u_j)$, the more likely the two users have collusion.
For each user $u_i$, we first calculate the $Score_{F}$ between user $u_i$ and all other users,
and then the spamicity score of user $u_i$ is computed by averaging the sum of 
\emph{Frobenius-distance based} (FD) scores of $n$ users, whose FD scores are ranked in the top-$n$ list of $Score_{F}(u_i,*)$.

\subsection{Optimization}
\label{sec-opt}

In this section, we present the solution to the optimization problem stated in Eq. (\ref{con:loss}), which is optimized as follows,

%
%\textbf{Step I}:
\paratitle{Optimize $\mathcal{L}_{d}$}.
We first focus on the loss function of $\mathcal{L}_{d}$ and it can be rephrased as follow:
{
\vspace{0pt}
\small
\begin{equation}
\begin{split}
\mathcal{L}_{d}
&= \sum_{(i, j) \in E} \matrix{I}_{ij}w_{ij}\|\vec{u}_{i}-\vec{u}_{j}\|_{F}^{2} \\
&= 2 tr(\matrix{U}^{T} \matrix{L} \matrix{U})
\label{con:lowOpti}
\end{split}
\end{equation}
}
where $\matrix{L} = \matrix{D} - \widetilde{\matrix{W}}$, and $\widetilde{\matrix{W}} = \matrix{I} \odot \matrix{W}$, $\matrix{D} \in \mathbb{R}^{|\matrix{U}| \times |\matrix{U}|}$ is a diagonal matrix and $\matrix{D}_{i,i}= \sum_{j} \widetilde{\matrix{W}}_{i,j}$,
$\matrix{W}$ is the weighted user-user matrix.
%~(\rf Section \ref{sec-direct}).
%
For each iteration, we first update $\matrix{I}$ and then update $\matrix{L}$.

According to the Eq.(\ref{con:lowOpti}), we can utilize SGD to minimize $\mathcal{L}_{d}$,
and the partial derivatives of $\mathcal{L}_{d}$ with respect to $\matrix{U}$ can be computed by,
{
\vspace{0pt}
\small
\begin{equation}
\begin{split}
\frac{\partial \mathcal{L}_{low}}{\partial \matrix{U}} =  2 (\matrix{L} + \matrix{L}^{T}) \cdot \matrix{U}
\label{con:lowGradient}
\end{split}
\end{equation}
}
% where $\lambda$ is the learning rate.

%\textbf{Step II:}
\paratitle{Optimize $\mathcal{L}_{id}$}.
Here, we first calculate the partial derivative of $\mathcal{L}_{id}$  with respect to user $u_i$,
namely,
%which is defined by
%We then pay attention to node $u_{i}$ and treat it as center node. In detail, given the center node $u_{i}$ and its context node $u_{c}$, we update its embedding vector $\boldsymbol{u}_{i}$ based on Eq.(\ref{con:high}) as follows:
{
\vspace{-5pt}
\small
\small
\begin{equation}
\begin{split}
\frac{\partial \mathcal{L}_{id}}{\partial \vec{u}_{i}} = -  \sum_{z \in {u_{c}} \cup N^{\kappa}(u_{i})}  (\matrix{O}(z, u_{i})-\sigma(\vec{u}_{i}^{T}\matrix{\Phi}_{z}))\matrix{\Phi}_{z}
\label{con:highGradient1}
\end{split}
\end{equation}
}
where $\matrix{O}(z, u_{i})$ is an indicator function, $\matrix{O}(z, u_{i}) = 1$ if $z$ is the neighbor of $u_{i}$ and 0 otherwise;
$N^{\kappa}(u_{i})$ is $\kappa$ negative samples for $u_{i}$.
Then the  partial derivative of $\mathcal{L}_{id}$  with respect to user $u_i$
%when her role is a neighbor, which
is updated by
{
\vspace{0pt}
\small
\begin{equation}
\begin{split}
\frac{\partial \mathcal{L}_{id}}{\partial \matrix{\Phi}_{z}} =   - \sum_{z \in {u_{c}} \cup N^{\kappa}(u_{i})} (\matrix{O}(z, u_{i})-\sigma(\matrix{u}_{i}^{T}\matrix{\Phi}_{z})) \vec{u}_{i}
\label{con:highGradient2}
\end{split}
\end{equation}
}

Subsequently,
the regularization term can be updated by Eq. (\ref{con:regGradient}),
{
\vspace{0pt}
\small
\begin{equation}
\begin{split}
\frac{\partial \mathcal{L}_{high}}{\partial \matrix{U}} =  2 \matrix{U}.
\label{con:regGradient}
\end{split}
\end{equation}
}
{
\begin{algorithm}[!h]
    \caption{The Learning Process of COSD}
    \label{alg:Framwork}
    \LinesNumbered
    \KwIn{$\forall <u_{i}, p_{j}, r_{ij}, t_{ij}, c_{j}>$}
    \KwOut{$\boldsymbol{U}$, $S$: User embedding matrix and spamicity scores list}
    Construct Signed Network based on Eq. (\ref{eq:weight})(\ref{eq:loss_dir_rel}) \\
    Obtain walk sequences by RandomWalk. \\
    \While{not converged}{
        Compute $\nabla \mathcal{L}_{low}$ based on Eq. (\ref{con:lowGradient})  \\
    	Compute $\nabla \mathcal{L}_{reg}$ based on Eq.  (\ref{con:regGradient}) \\
    	\For {$u_{i} \in U$}{
     		Sample $\kappa$ negative nodes \\
    		Compute $\nabla \mathcal{L}_{high} $based on Eq. (\ref{con:highGradient1}) (\ref{con:highGradient2})
    	}
    	Update $\boldsymbol{U}$ and $\boldsymbol{\Phi}$ \\
    }
    \For {$u_{i} \in U$} {
    	\For {$u_{j} \in U$} {
    		$r_{i, j} = \exp(-\|u_{i}-u_{j}\|_{F}^{2})$ \\
    		Insert ${r_{i, j}}$ into $R_{i}$
    	}
      	Accumulate top k maximum values in $R_{i}$ to $s_{i}$ \\
    	Insert $(s_{i}, u_{i})$ into S \\
    }
    RETURN $\boldsymbol{U}$, S
\end{algorithm}

For ease of reference, the learning process of our proposed method COSD
is summarized in Algorithm \ref{alg:Framwork}.
%
%we present the proposed procedure of COSD (Algorithm \ref{alg:Framwork}) in detail.
Specifically,
%detail the proposed procedure of COSD in Algorithm \ref{alg:Framwork}.
it first builds the user-based signed network and generates the walk sequences using Eq. (\ref{eq:weight})
and Eq. (\ref{eq:loss_dir_rel}),
and then iteratively update the user embedding matrix $\boldsymbol{U}$ and $\boldsymbol{\Phi}$
via Eq.(\ref{con:lowGradient})-Eq.(\ref{con:regGradient}).
Finally, the spamicity scores are calculated via Frobenius Distance and output the spamicity score list $S$ and the user embedding matrix $\boldsymbol{U}$.

\paratitle{Time Complexity}.
The procedure of our proposed method consists of the following
sub-components, the computation costs of which are,
(i) The construction of the user-based signed network ($O(|\matrix{U}|^2)$);
(ii) The iteration of updating the user embedding matrix, $O(\matrix{I} \times |\matrix{U}| \times \kappa \times K)$,
where $I$ is the number of iterations, $\kappa$ is the number of negative sampling, $K$ is the dimension of the representation vector;
and (iii) The calculation of the spamicity scores list ($O(|\matrix{U}|^2 \times log(|\matrix{U}|))$).
Therefore, the time complexity of our approach is 
$O(I \times |\matrix{U}| \times \kappa \times K)$,
as  $O(I \times \kappa \times K)>O(|\matrix{U}| \times log ( |\matrix{U}|))$.

%Firstly, it requires $O(|\matrix{U}|^2)$ to calculate the pairwise features for constructing user-based signed network. Then, the cost of iteratively updating the user embedding matrix is $O(\matrix{I} \times |\matrix{U}| \times \kappa \times K)$, where $I$ is the number of iterations, $\kappa$ is the number of negative sampling, $K$ is the dimension of the representation vector. Further more,
%the method requires $O(|\matrix{U}|^2 \times log ( |\matrix{U}|) )$ complexity to calculate spamicity scores list. 
%As $O(I \times \kappa \times K)$ always greater than $O(|\matrix{U}| \times log ( |\matrix{U}|))$ during the iterative process, the final time complexity is $O(I \times |\matrix{U}| \times \kappa \times K)$.

}

\section{EXPERIMENTS}

\begin{table*}[t]
\centering
 \caption{Statistics of the used datasets.}
  \label{tab:freq}
    \begin{tabular}{l|l|l|l}
      \hline
        \multicolumn{2}{c|}{} & \textbf{AmazonCn} & \textbf{YelpHotel}\\
        \hline
        \multicolumn{2}{c|}{ \textbf{\# reviews} }  & 1,205,125 & 688,313 \\
        %\hline
        \multicolumn{2}{c|}{ \textbf{\# reviewers} }  & 645,072 & 5,122 \\
        %\hline
        \multicolumn{2}{c|}{ \textbf{\# products} }  & 136,785 & 282,974 \\
        \hline
        \multirow{2}{*}{\textbf{\# labeled data}} & \textbf{\# colluders} & 1,937 & 450 \\ \cline{2-4}
        & \textbf{\# non-colluders} & 3,118 & 672 \\
      \hline
    \end{tabular}
\end{table*}

\subsection{Data Sets}
\label{sec-dataset}
%
%We use two public real-world datasets for evaluation, \ie \emph{AmazonCn}, \emph{YelpHotel}.
%We compared the proposed methods with the state-of-the-art methods on two public
%Our experiments are conducted on the following two datasets.
%f

In this section, we use two public real-world datasets for evaluation, \ie \emph{AmazonCn}, \emph{YelpHotel},
as compared to the state-of-the-art methods.
The statistics of the datasets are summarized in Table \ref{tab:freq}:
(i) \textbf{AmazonCn}. is a collection of real consumers' reviews from \emph{Amazon.cn}~\cite{xu2013uncovering}, which already has \emph{gold} standard of collective spammers whose reviews are filtered by \emph{AmazonCn}, and then gather those tagged users with their co-rating behaviors;
and (ii) \textbf{YelpHotel}. contains real reviews about hotel on \emph{Yelp.com}, which is collected by Yelp's own filtering mechanism~\cite{mukherjee2013yelp}.

Note that there are not labels of collective spammer in YelpHotel, and thus we follow the work \cite{mukherjee2012spotting}
to form spammer groups via mining user common behaviors with \emph{Frequent Itemset Mining}~(FIM).
%find the forming spammer groups by mining user common behaviors using \emph{Frequent Itemset Mining}~(FIM).}
%to build spammer groups via mining user common behaviors.}
%
%As it has no collective spammers, we follow the \cite{mukherjee2012spotting} to utilize \emph{Frequent Itemset Mining}~(FIM) for extracting users with common behaviors to be formed as groups.

\subsection{Evaluation Metrics}
As mentioned, the output of collective opinion spammer detection problem is a ranking list of candidates with the spamicity scores, which
can be deemed to the likelihood of candidates participating in opinion spam.
As such, to evaluate the detection performance of different approaches,
we follow the work in \cite{xu2017online} to adopt four well-known metrics,
\ie
(i) \textbf{Average Precision~(AP)}, which measures the average precision of collective spammer retrieving over the interval of $r$ (recall value) from 0 to 1;
%calculates the average precision of spammer retrieving over the interval of $r$ from $0$ to $1$, where $r$ denotes recall value;
%
(ii) \textbf{Area Under ROC Curve~(AUC)}, which measures the accuracy based on \emph{False Positive Ratio}~(FPR) against \emph{True Positive Ratio}~(TPR) for binary classification;
(iii) \textbf{Precision@k~(P@k)},~which measures the percentage of true spammers in the top-K returned candidates;
and
(iv) \textbf{Normalized Discounted Cumulative Gain@k~(NDCG@k)}, the performance of the detection model based on the ground-truth (\ie spammer (1)/non-spammer (0)) of the returned spammers, which is the normalization of Discounted Cumulative Gain (DCG) at each position for a chosen value of k.

\subsection{Baseline Methods}
We compare our model with the following state-of-the-art spam detection methods:

\paratitle{GsRank}.
\cite{mukherjee2012spotting}: This method employs an ranking model for spotting spammer groups by extracting group behavior features and individual behavior features
over user-product-group relationships, which is iteratively computed for the spamicity scores of users in the reviewer group.

\paratitle{FraudInformer}.
\cite{xu2015combating}: This method extends the Markov Random Walk model to obtain a ranking of reviewers' spamicity scores by exploring multiple heterogenous pairwise features from reviewers' rating behaviors and linguistic patterns.
%An unsupervised and intuitive colluder detecting method which explore multiple heterogenous pairwise features from reviewers' rating behavior and linguistic pattern, it extends the Markov Random Walk model to obtain a ranking of reviewers' spamicity.

\paratitle{HoloScope}.
%focus on graph topology and temporal information for detecting fraud users using a scalable dense block detection method.
%
\cite{liu2017holoscope}: This method detects fraud users by using a scalable dense block detection method to explore graph topology and temporal spikes.
%A scalability dense block detection method that uses information from graph topology and temporal spikes to find groups of suspicious users. And it also makes holistic use of several signals and rating deviation in framework.

\paratitle{FRAUDSCAN}.
\cite{xu2017online}:  This method utilizes the regularized matrix factorization to learn the fraud campaign embedding for detecting fraud campaign.
%
%A powerful principled optimization model for detecting fraud campaign, which utilize regularized matrix factorization for fraud campaign embedding.

\paratitle{FraudNE}.
\cite{zheng2018fraudne}:
It is a deep structure embedding framework
to capture the highly non-linear structure information
between the vertexes,
the different types of which are jointly embedded into a latent space
via multiple non-linear layers.
The spammers are detected through a clustering algorithm
by averaging the degrees of each cluster to find a high-density area in the latent space.

\paratitle{ColluEagle}.
\cite{Wang2020}:
A powerful Markov random field (MRF)-based model which builds the user network over co-review behaviors, and then employs the loopy belief propagation model to calculate the spamicity scores of users under a pairwise based Markov Random Field (MRF) framework.
%detects collusive review by considering both network effects and time effects.

\paratitle{Our method}. Our proposed collective opinion spammer detection method is called \emph{COSD}, which simultaneously learn the \emph{direct relevance embeddings} and \emph{indirect relevance embeddings} over the co-reviewed correlations for detecting spammers.
To analyze the different impacts of the two types of relevance embeddings, we also evaluate two  variants of our model, that is,
(i) \textbf{\emph{COSD-D}}, this is a variant of our COSD model and we only optimize \emph{direct} relevance loss;
%In addition, in order to fully demonstrate the effectiveness of direct relevance section and indirect relevance section, we also evaluated two variants of our model:
and (ii) \textbf{\emph{COSD-ID}}, this is a variant of our COSD model that we only optimize \emph{indirect} relevance loss.

\begin{figure*}[t]
  \begin{center}
     \subfloat[{AmazonCn}]{\includegraphics[width=0.5\textwidth,angle=0]{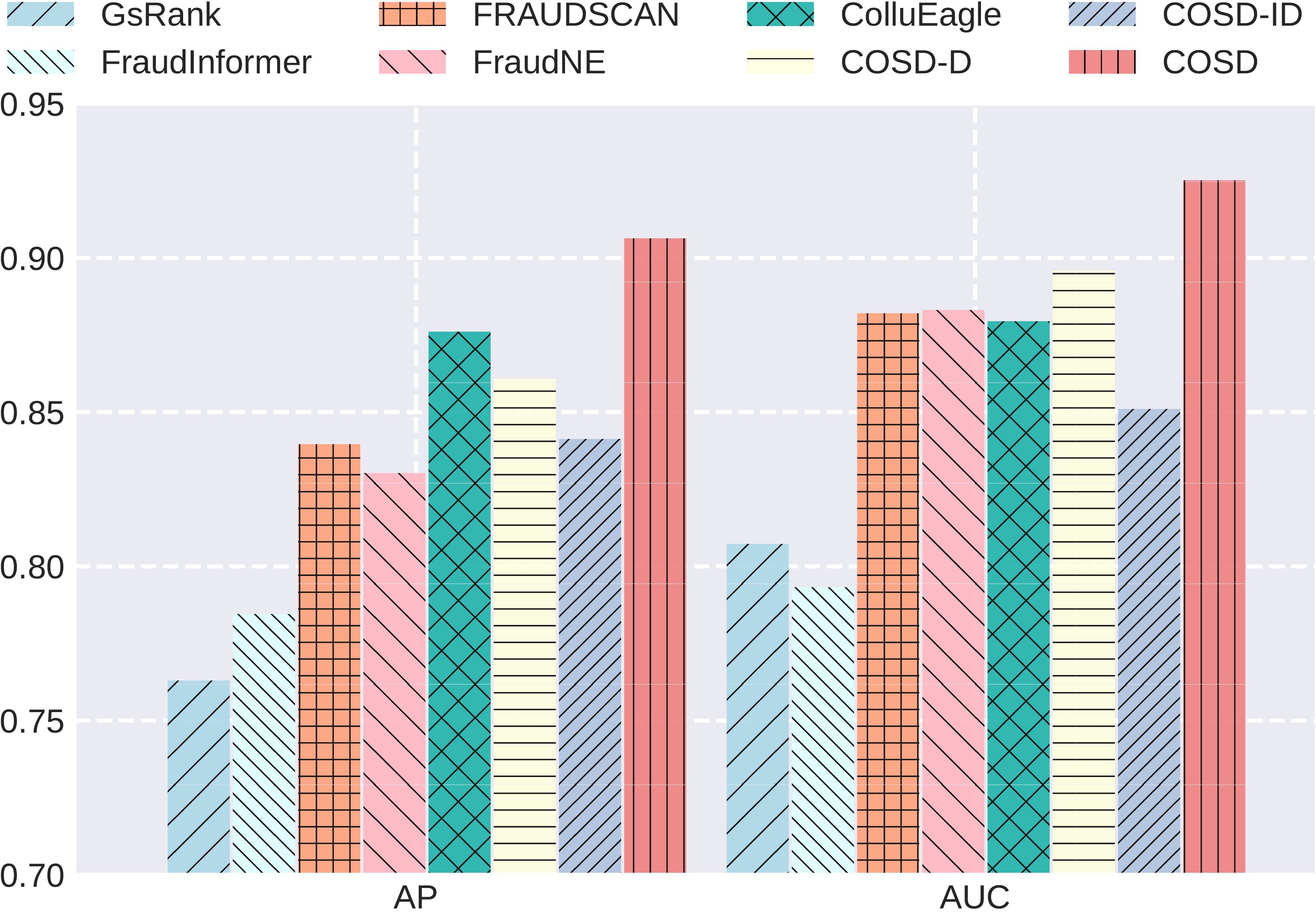}
     \label{fig:AmazonResults}
     }
     \subfloat[{YelpHotel}]{\includegraphics[width=0.5\textwidth, angle=0]{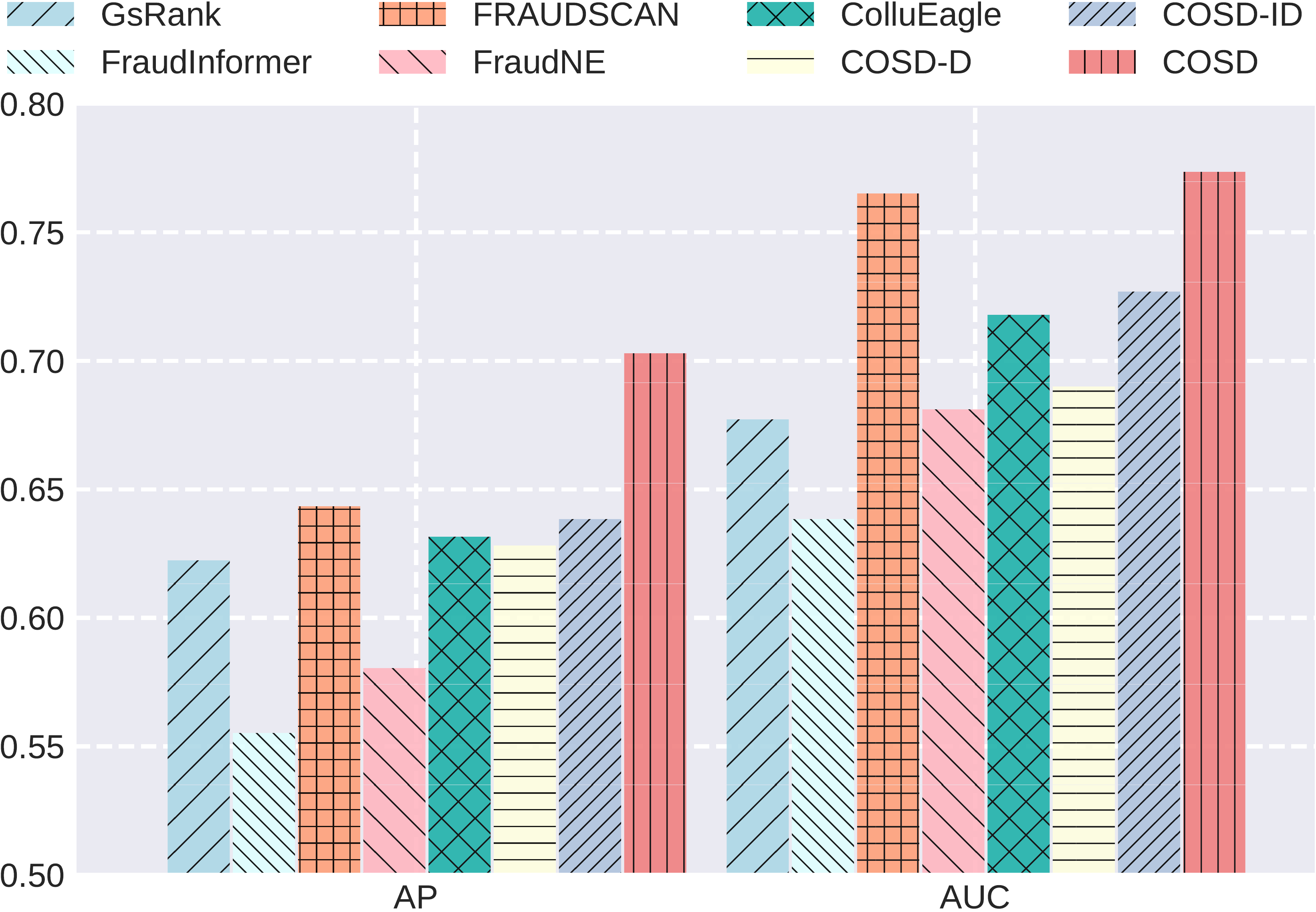}
     \label{fig:YelpResults}
     }
     %\vspace{-1ex}
     \vspace{-0.3cm}\caption{Effectiveness comparison between COSD and state-of-the-art approaches on the two datasets.}
     \label{fig:Results}
  \end{center}
%\vspace{0ex}\hrule
\vspace{-1cm}
\end{figure*}
%\paratitle{COSD-L}.
%%
%This is a variant of the COSD model and we only optimize direct relevance loss in this variant.
%
%\paratitle{COSD-H}.
%%
%Different from COSD-L, this is a variant of the COSD model that we only optimize indirect relevance loss in this variant.

\subsection{Implementation Details}
For \emph{GsRank}, we utilize FIM to extract candidate spammer groups as mentioned in~\cite{mukherjee2012spotting}.
For \emph{FraudInformer}, we treat all features fairly as mentioned in \cite{xu2015combating}, and the computation ends when $\epsilon=10^{-6}$.
%when the difference between two iterations is less than $10^{-6}$, we terminate the algorithm.
%
For \emph{HoloScope}, we use the implementation provided by the shared source code\footnote{https://github.com/shenghua-liu/HoloScope}.
% and as it is a dense block detection method, we only compare our model with it in terms of P@k.
For \emph{FraudInformer}, we keep the hyper-parameters~(\ie $\alpha_1, \alpha_2, \alpha_3, k$) consistent with the original paper and conduct grid search for the learning rate in $\{ 0.1, 0.01, 0.001, 0.0001 \}$.
For \emph{FraudNE}, the learning rate is searched in $\{ 0.01, 0.001, 0.0001\}$, the trade-off parameters $\alpha$ is searched in $\{ 0.1, 0.5, 1, 2\}$ and the scale factor $\beta$ is searched in $\{ 5, 10, 15, 20\}$.
For \emph{ColluEagle}, we conduct grid search for the variance of a normal distribution $\sigma_1, \sigma_2$ in $\{ 1, 10, 30, 60, 90, 120 \}$ and the prior rate in $\{0, 0.2, 0.4, 0.6, 0.8\}$.

%The parameters for our proposed method are empirically set as follows:
%
The parameters for our proposed method are empirically set as follows:
The embedding size $K$ is set as $64$ on AmazonCn and $128$ on YelpHotel, respectively.
For the setting of the random-walk of each node, $\gamma=30$, $t=8$, $\omega=5$, $\kappa=8$ for two datasets;
and for $\{\beta, n\}$, we set is as $\{0.6, 25\}$ for \emph{AmazonCn},
and $\{0.4, 40\}$ for \emph{YelpHotel}.
%
%the number of walks per node $\gamma$ is set as $30$, and the walk length $t$  as $8$, window size $\omega$ as $5$ and negative samples $\kappa$ as $8$ both on Amazon and YelpHotel. And as for $\{ \alpha, n \}$, we set $\{ 0.6, 25 \}$ on AmazonCn and $\{0.4, 40\}$ on YelpHotel.

\subsection{Evaluation Results and Analysis}

\paratitle{Comparison of the Spammer detection Performance}.
This experiment is to evaluate the effectiveness of identifying the spammers by our approach with the baseline methods on \emph{AmazonCn} and \emph{YelpHotel}.
Figure (\ref{fig:AmazonResults}) and (\ref{fig:YelpResults}) shows \emph{AP} and \emph{AUC} of each method on such two datasets.
Each method is performed $10$ times, respectively.

%The experiment results of various methods on AmazonCn and YelpHotel are presented in Figure (\ref{fig:AmazonResults}) and (\ref{fig:YelpResults}). We show the results in terms of AP and AUC on two real-world datasets. All of the results in Figure \ref{fig:Results} are the average over 10 runs.

From Figure (\ref{fig:AmazonResults}) and (\ref{fig:YelpResults}), we can observe that: \emph{COSD} consistently outperforms all baselines on all metrics.
For example, \emph{COSD} outperforms \emph{GsRank} by [$18.8\%, 14.5\%$] in terms of \emph{AP} and \emph{AUC}  on Amazon and [$12.9\%, 1.6\%$] on YelpHotel, respectively.
% First, COSD performs better than GsRank on all metrics. In detail, COSD outperforms GsRank by [$18.8\%, 14.5\%$] in terms of AP and AUC respectively on Amazon and [$12.9\%, 1.6\%$] on YelpHotel.
This is because \emph{COSD} is capable of learning the information from  pairwise user relations with behavioral information,
while GsRank is based on the group information, which might omit several subtle effective information from users.
%
%This is because COSD analyzes the relation between pairwise user with behavioral information while GsRank is based on group, which will omit many effective information.
In addition, \emph{COSD} outperforms \emph{FraudInformer} by [$17.9\%, 16.6\%$] on Amazon and [$26.5\%, 21.8\%$] on YelpHotel in terms of AP and AUC, respectively.
%
%Second, our model outperforms FraudInformer by [$17.9\%, 16.6\%$] in terms of AP and AUC respectively on Amazon and [$26.5\%, 21.8\%$] on YelpHotel.
However, \emph{COSD} and \emph{FraudInformer} both make use of pairwise user features for collective spammers detection, which demonstrate that
\emph{COSD} is more effective and robust in exploiting the relationships between users from different-levels, \ie \emph{direct relevance} and \emph{indirect relevance},
%Significantly, both COSD and FraudInformer utilize pairwise user features to detect collective spammers, it demonstrates that our model is more effective and robust by exploring the relationship between users from different levels.
which can effectively distinguish users with low likelihood of collusion and thereby avoiding heaps of noise.
%Moreover, it also shows that embedding approaches is capable of learning better from features for colluder detection.
%And it clearly shows that on both datasets, our model performs better than all the other methods in terms of AP and AUC.
%
Comparing with \emph{FraudNE}, our method outperforms it on both datasets, as \emph{FraudNE} neglects the rating inforamtion and time information in the datasets, besides it is hard for \emph{FraudNE} to capture the potential collusion between pair of users.
Additionally, we also observe that \emph{COSD} outperforms \emph{FRAUDSCAN} by [$7.9\%, 9.2\%$]  on \emph{AmazonCn},
and [$4.8\%, 1.6\%$] on \emph{YelpHotel} in terms of \emph{AUC} and \emph{AP}, respectively.
And comparing with state-of-the-art method ColluEage, the performance is improved by [$3.4\%, 5.1\%$] on on \emph{AmazonCn}, and [$11.3\%, 7.7\%$] on \emph{YelpHotel} in terms of \emph{AUC} and \emph{AP}, respectively.
%
%By comparing COSD with the start-of-the-art methods FRAUDSCAN, the performance is improved by $7.9\%$ in terms of AP and $9.2\%$ in terms of AUC on AmazonCn, and $4.8\%$ in terms of AP and $1.6\%$ in terms of AUC on YelpHotel.
The reason might be \emph{COSD} effectively captures the possibility of collusion for pairwise users who do not have common co-reviewed records,
and \emph{COSD} can model users' neighbor structure over the \emph{indirect relations} for detection.

%We believe that one of the reasons for the improvement is that we effectively captured the possibility of collusion between two users who do not have a common review record.

\begin{figure}[t]
   \begin{center}
        \subfloat[{Precision@k on AmazonCn}]{\includegraphics[width=0.5\linewidth,
        angle=0]{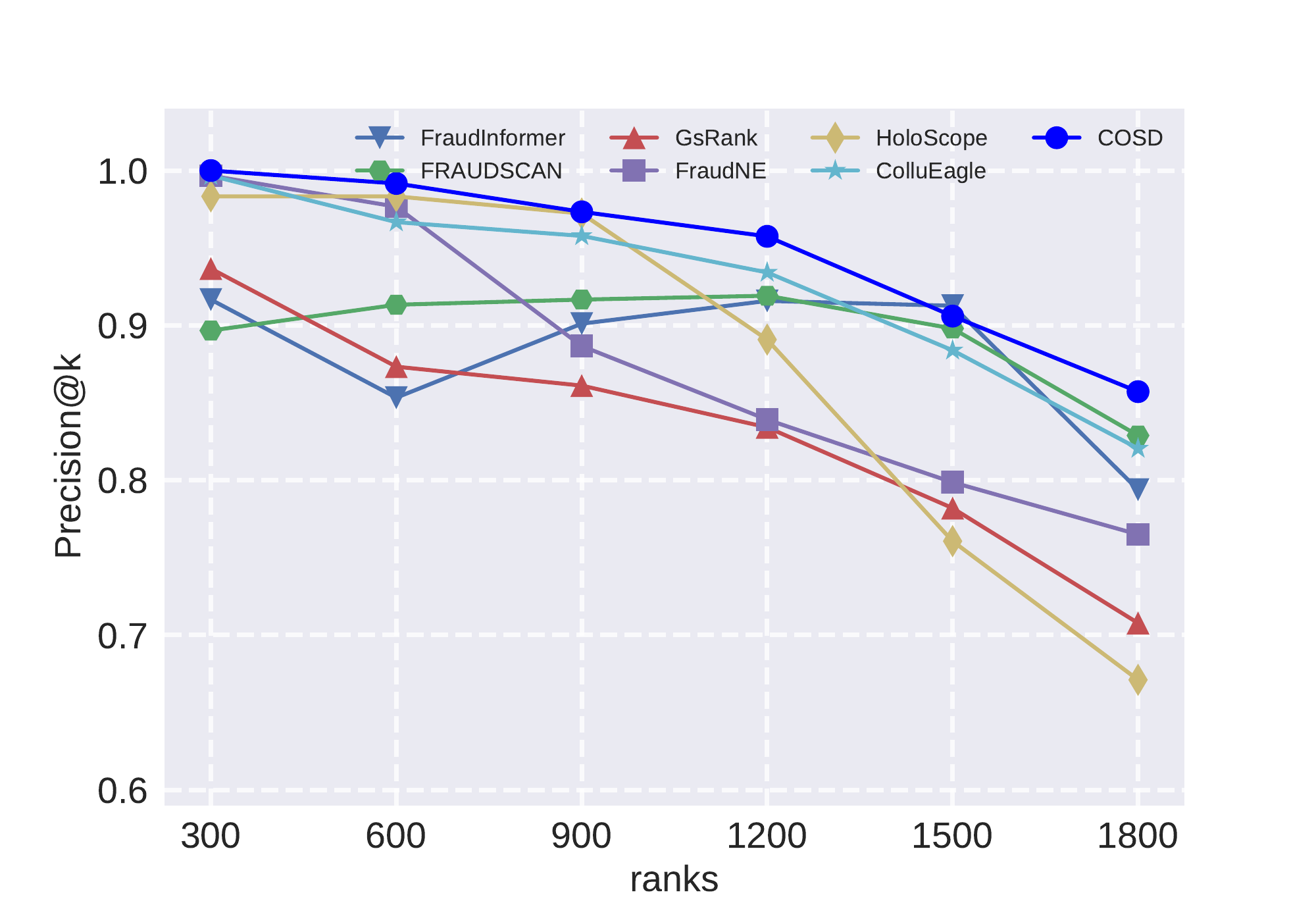}}
        \subfloat[{Precision@k on YelpHotel}]{\includegraphics[width=0.5\linewidth, angle=0]{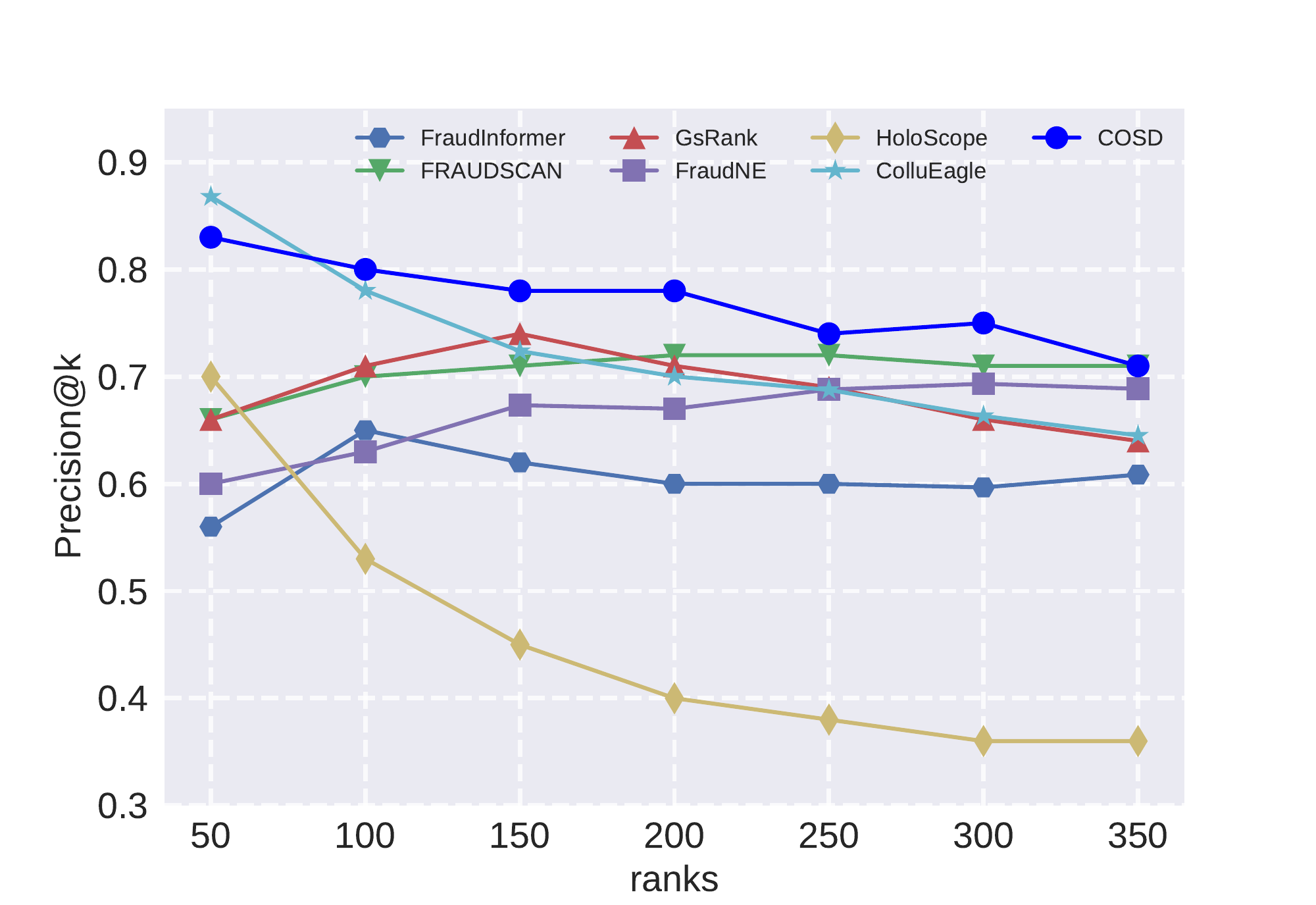}}

        \subfloat[{NDCG@k on AmazonCn}]{\includegraphics[width=0.5\linewidth,
        angle=0]{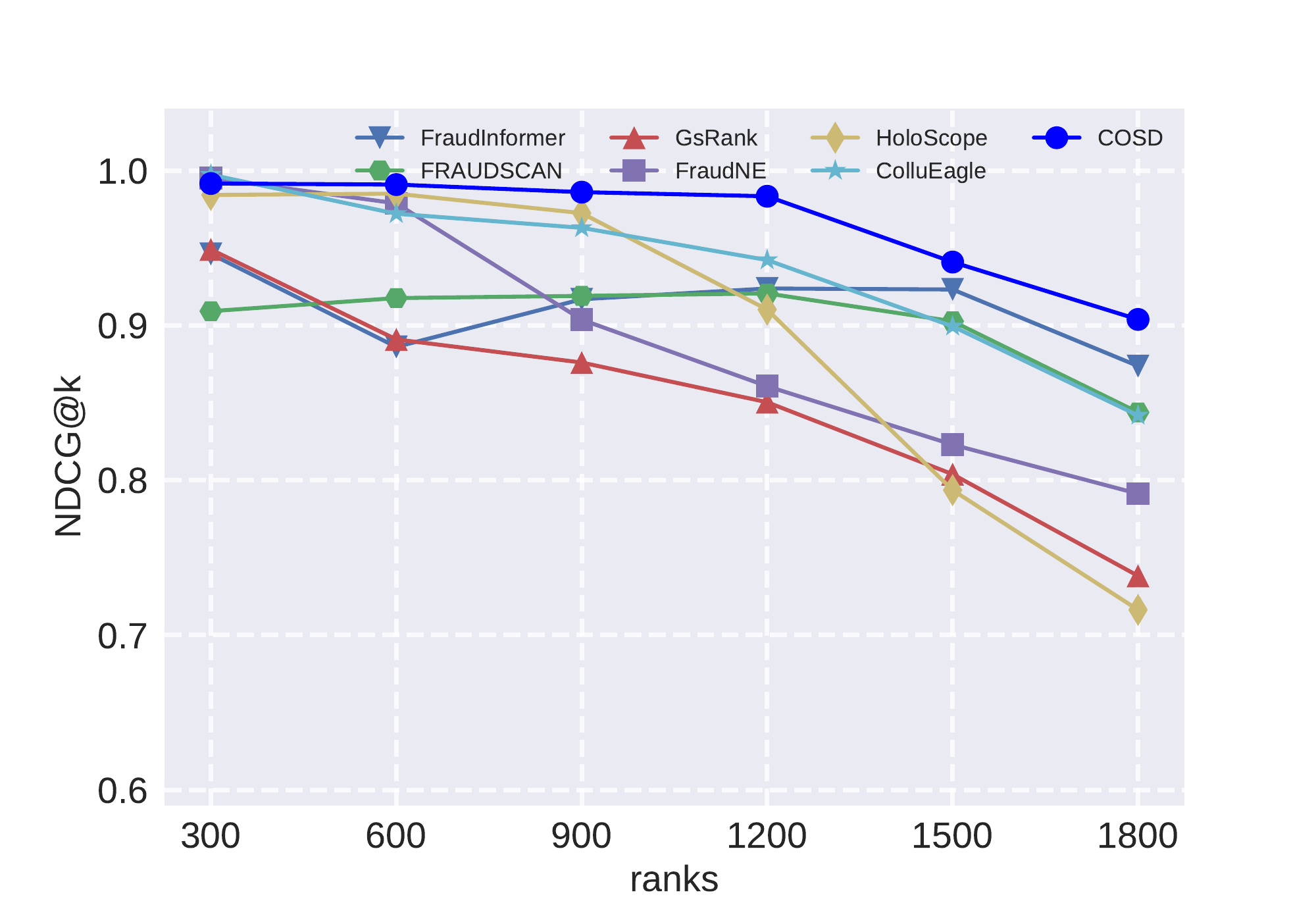}}
        \subfloat[{NDCG@k on YelpHotel}]{\includegraphics[width=0.5\linewidth, angle=0]{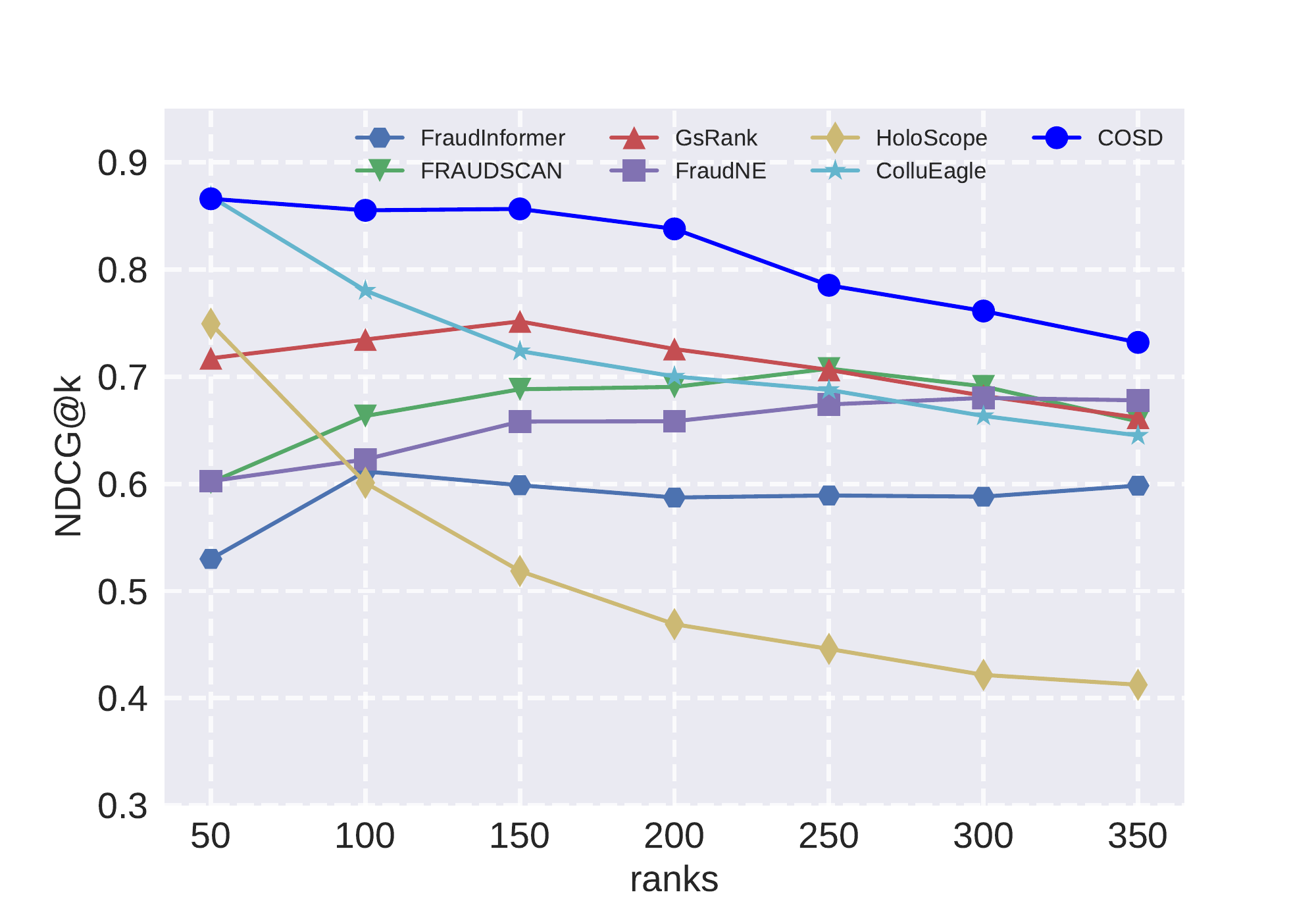}}
        \caption{Model performance in terms of Precision@k.}
      \label{fig:Results_Precision}
   \end{center}
%\vspace{0ex}\hrule
 \vspace{0cm}
\end{figure}

From Figure (\ref{fig:Results_Precision}) we can observe that most of methods perform worse when rank $k$ increases, and our method outperforms other methods consistently on AmazonCn and Yelp datasets in terms of Precision@k and NDCG@k, which also demonstrates the effectiveness of our method.
More importantly, \emph{COSD} can combine the\emph{ direct relevance} and \emph{indirect relevance} over user behavioral features
for jointly optimizing the losses of learning such two relevance embedding, and thus the learnt user embeddings might  preserve not only the behavioral information but also the topology information, which makes the users with \emph{direct}  and \emph{indirect} collusive connections close in the learnt low-dimensional vector space.

\paratitle{The Impact of Direct Relevance Embedding}.
\label{sec:dre}
From Figure (\ref{fig:Results}), we can observe that \emph{COSD-D} and \emph{COSD-ID} achieve good results.
\emph{COSD-D} outperforms all baselines on \emph{AmazonCn} in terms of \emph{AP},
which demonstrate the effectiveness of using a signed network with \emph{positive} and \emph{negative} links,
and it also shows that optimizing the \emph{direct relevance} is capable of more accurately reflecting the realistic relationships in low-dimensional space.

%On AmazonCn COSD-L demonstrate significant improvement and it outperforms all other methods in terms of AP. On the one hand it shows that it is meaningful to divide the relation between users into positive and negative, on the other hand it shows the effectiveness of the signed network in dealing with relational problems. We obtain a more realistic relationship in low-dimensional space by optimizing the direct relevance. It demonstrates that low order proximity reflects the relationship between users more accurately.

\paratitle{The Impact of indirect Relevance}.
\label{sec:idre}
From Figure (\ref{fig:Results}), we can observe that \emph{COSD-ID} does not perform as well as \emph{COSD-D}, however it still outperforms other baselines on \emph{AmazonCn} in terms of \emph{AP}, which demonstrates that the effectiveness of performing a random walk along with the positive links on a built user-based signed network while discarding the negative links with weak collusion signals (even noise) of being colluders, and thus it can focus on the analysis of the filtered neighbor structure for more accurately capturing the collusion signals, and it also shows that using the indirect relations is useful for exploring the implicit associations
for any pairwise uses in these fraud campaigns.

%We can get the following conclusion from results: 1) signed networks is effective in distinguishing user relationships. It can better express the connection between users. 2) Using the user's neighbor information is an effective way to detect group spammers, which can find potential connections between two users. 3) Simultaneous use of direct relevance and indirect relevance can more effectively represent the true connection between users in low-dimensional representations, which can simultaneously maintain direct and indirect connections of users.

\begin{figure}[!t]
   \begin{center}
      \subfloat[{sensitivity of $\beta$}]{\includegraphics[width=.5\linewidth, angle=0]{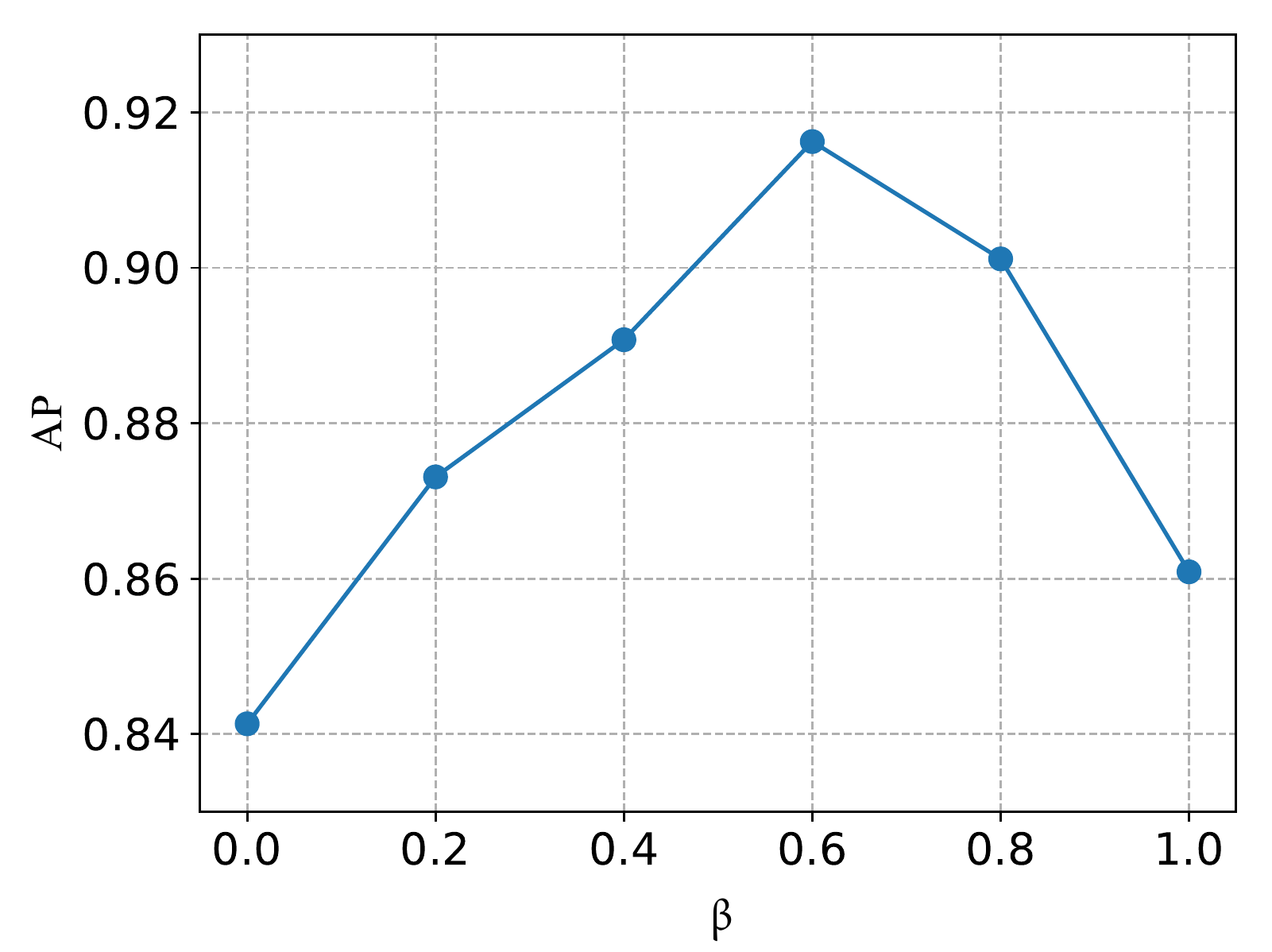}
      \label{fig:sensitive_alpha}}
      \subfloat[{sensitivity of $\zeta$}]{\includegraphics[width=.5\linewidth, angle=0]{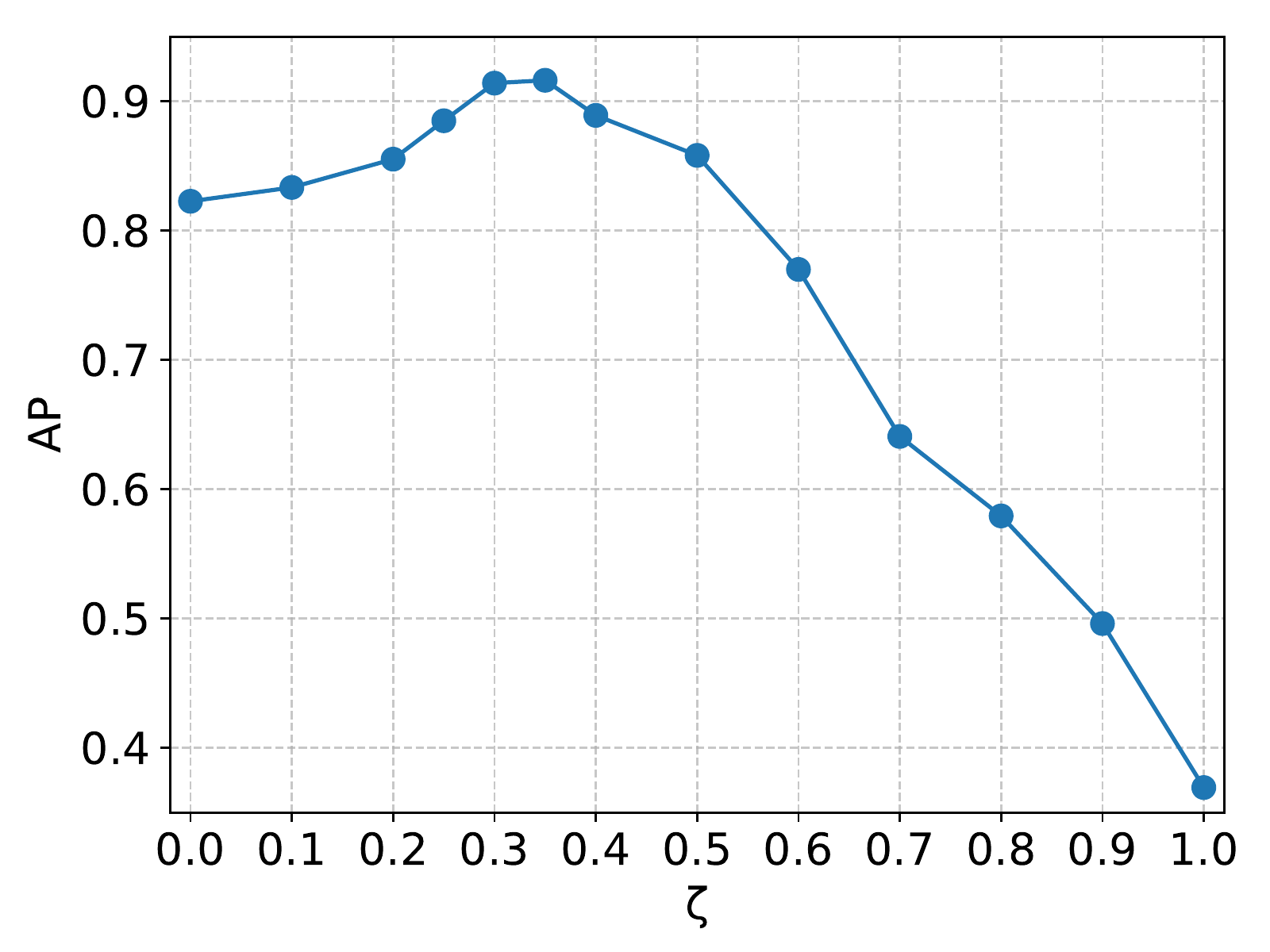}
      \label{fig:sensitive_zeta}}\\
      \subfloat[{sensitivity of $K$}]{\includegraphics[width=.5\linewidth, angle=0]{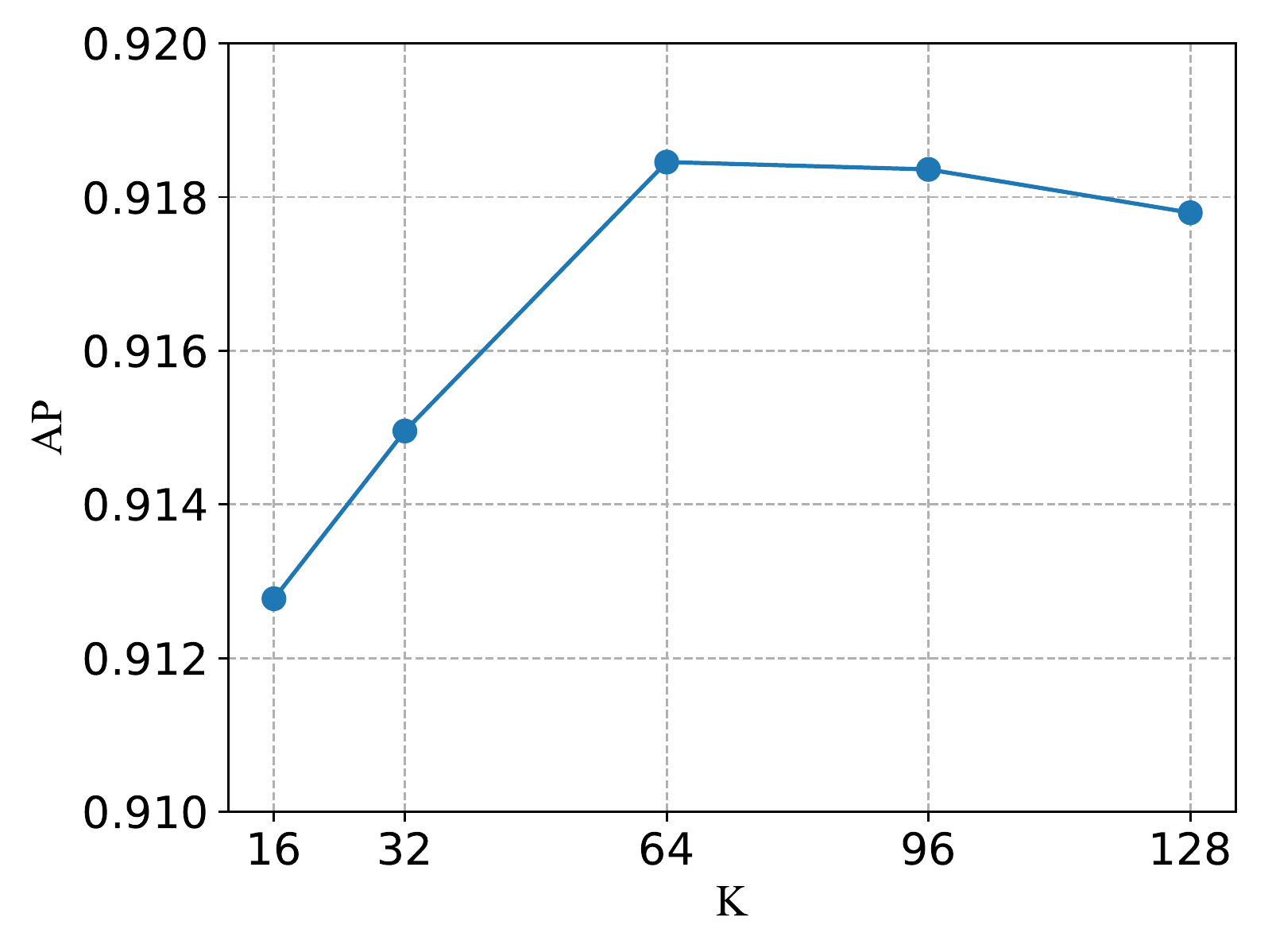}
      \label{fig:sensitive_K}}
      %\subfloat[{sensitivity of $n$}]{\includegraphics[width=.5\linewidth, angle=0]{picture/sensitive_k.pdf}
%      \label{fig:sensitive_k}}
\vspace{0cm}
      \caption{Illustration of the effect of different parameters, \ie $\alpha$ and $\zeta$
      % and $K$
      on \emph{AmazonCn}.}
   \end{center}
%\vspace{0ex}\hrule
 \vspace{0cm}
\end{figure}

\subsection{On the Sensitivity of Parameters}
In this section, we study the impact of three parameters, \ie $\beta, \zeta, K$, in our method\footnote{We only perform parameter sensitivity experiment on AmazonCn dataset as the scale of the YelpHotel dataset is too small to observe significant changes.}.

For parameter $\beta$ (used in Eq. \eqref{con:loss}), which controls the contributions of  \emph{direct relevance embeddings} and \emph{indirect relevance embeddings} in our model, and
%We can observe that
As shown in Figure (\ref{fig:sensitive_alpha}),
our proposed method achieves the best result when $\beta=0.6$, and simultaneously making use of both
\emph{direct relevance embedding} and \emph{indirect relevance embedding}
outperforms the extreme cases when only considering the \emph{indirect relevance embedding} ($\beta=0$) or the \emph{direct relevance embedding} ($\beta=1$),
which is consistent with the former analysis, \ie \emph{direct relevance embedding} is more important than \emph{indirect relevance embedding}.

For parameter $\zeta$ (used in Eq. \eqref{eq:weight}), which is a thresholds to distinguish \emph{high}-probability and \emph{low}-probability collusion signal.
As shown in Figure (\ref{fig:sensitive_zeta}), our proposed method achieves the best result within the range of $[0.3,0.4]$,
which demonstrate that when the value of $\zeta$ is small, it might result in more noise involved,
while a large one would lead to the insufficiently learning of \emph{indirect relevance embedding} due to neglecting more indirect collusive relationships.

For parameter $K$, which is the dimensional (same for \emph{direct} and \emph{indirect}) of the representation embedding vector. As shown in Figure (\ref{fig:sensitive_K}), when varying  $K$, the performance first increases and then decreases, and the best perform is achieved when $K=64$, which demonstrate that our proposed method can well work with a low dimension vector space.

\section{Conclusion}

In this paper, we propose a novel \emph{unsupervised} network embedding-based approach to jointly combine direct and indirect neighborhood exploration for learning the user embeddings for more accurately identifying spam reviewers.
Experiments on two real-world datasets demonstrate that our proposed approach significantly outperforms all of baselines on all metrics, by learning the user representation via jointly optimizing the \emph{direct relevance} and \emph{indirect relevance}.

For future work, we plan to investigate the following issues:
1) trying to incorporate more pairwise features and prior knowledge information into the model to enhance the representation ability of direct relevance modeling;
%more robust pairwise features and prior knowledge information can be incorporated to enhance the ability of direct relevance modeling
%
%other forms of objective functions, such as semisupervised HITS; 
2) considering an unsupervised graph neural network algorithm to enhance the representation ability of user embedding,
via modeling the indirect relevance of users in the latent space;
and 3) conducting more comprehensive studies to investigate the sensitivity of the parameters in the
model.

\section*{Acknowledgments}
This work was supported in part by the National Natural Science Foundation of China under Grant No.61602197 and Grant No.61772076, and in part by Equipment Pre-Research Fund for The 13th Five-year Plan under Grant No.41412050801.

\bibliography{mybibfile}

\begin{thebibliography}{10}
\expandafter\ifx\csname url\endcsname\relax
  \def\url#1{\texttt{#1}}\fi
\expandafter\ifx\csname urlprefix\endcsname\relax\def\urlprefix{URL }\fi
\expandafter\ifx\csname href\endcsname\relax
  \def\href#1#2{#2} \def\path#1{#1}\fi

\bibitem{fornaciari2014identifying}
T.~Fornaciari, M.~Poesio, Identifying fake amazon reviews as learning from
  crowds.

\bibitem{ren2016deceptive}
Y.~Ren, Y.~Zhang, Deceptive opinion spam detection using neural network, in:
  Proceedings of COLING 2016, the 26th International Conference on
  Computational Linguistics: Technical Papers, 2016, pp. 140--150.

\bibitem{wang2016learning}
X.~Wang, K.~Liu, S.~He, J.~Zhao, Learning to represent review with tensor
  decomposition for spam detection, in: Proceedings of the 2016 Conference on
  Empirical Methods in Natural Language Processing, 2016, pp. 866--875.

\bibitem{li2017bimodal}
H.~Li, G.~Fei, S.~Wang, B.~Liu, W.~Shao, A.~Mukherjee, J.~Shao, Bimodal
  distribution and co-bursting in review spam detection, in: Proceedings of the
  26th International Conference on World Wide Web, 2017, pp. 1063--1072.

\bibitem{ott2011finding}
M.~Ott, Y.~Choi, C.~Cardie, J.~T. Hancock, Finding deceptive opinion spam by
  any stretch of the imagination, arXiv preprint arXiv:1107.4557.

\bibitem{ye2015discovering}
J.~Ye, L.~Akoglu, Discovering opinion spammer groups by network footprints, in:
  Joint European Conference on Machine Learning and Knowledge Discovery in
  Databases, Springer, 2015, pp. 267--282.

\bibitem{mukherjee2013yelp}
A.~Mukherjee, V.~Venkataraman, B.~Liu, N.~Glance, What yelp fake review filter
  might be doing?, in: Seventh international AAAI conference on weblogs and
  social media, 2013.

\bibitem{lim2010detecting}
E.-P. Lim, V.-A. Nguyen, N.~Jindal, B.~Liu, H.~W. Lauw, Detecting product
  review spammers using rating behaviors, in: Proceedings of the 19th ACM
  international conference on Information and knowledge management, 2010, pp.
  939--948.

\bibitem{jindal2010finding}
N.~Jindal, B.~Liu, E.-P. Lim, Finding unusual review patterns using unexpected
  rules, in: Proceedings of the 19th ACM international conference on
  Information and knowledge management, 2010, pp. 1549--1552.

\bibitem{li2011learning}
F.~H. Li, M.~Huang, Y.~Yang, X.~Zhu, Learning to identify review spam, in:
  Twenty-second international joint conference on artificial intelligence,
  2011.

\bibitem{xie2012review}
S.~Xie, G.~Wang, S.~Lin, P.~S. Yu, Review spam detection via temporal pattern
  discovery, in: Proceedings of the 18th ACM SIGKDD international conference on
  Knowledge discovery and data mining, 2012, pp. 823--831.

\bibitem{gunnemann2014detecting}
S.~G{\"u}nnemann, N.~G{\"u}nnemann, C.~Faloutsos, Detecting anomalies in
  dynamic rating data: A robust probabilistic model for rating evolution, in:
  Proceedings of the 20th ACM SIGKDD international conference on Knowledge
  discovery and data mining, 2014, pp. 841--850.

\bibitem{gunnemann2014robust}
N.~G{\"u}nnemann, S.~G{\"u}nnemann, C.~Faloutsos, Robust multivariate
  autoregression for anomaly detection in dynamic product ratings, in:
  Proceedings of the 23rd international conference on World wide web, 2014, pp.
  361--372.

\bibitem{li2015analyzing}
H.~Li, Z.~Chen, A.~Mukherjee, B.~Liu, J.~Shao, Analyzing and detecting opinion
  spam on a large-scale dataset via temporal and spatial patterns, in: ninth
  international AAAI conference on web and social Media, 2015.

\bibitem{kc2016temporal}
S.~KC, A.~Mukherjee, On the temporal dynamics of opinion spamming: Case studies
  on yelp, in: Proceedings of the 25th International Conference on World Wide
  Web, 2016, pp. 369--379.

\bibitem{ye2016temporal}
J.~Ye, S.~Kumar, L.~Akoglu, Temporal opinion spam detection by multivariate
  indicative signals, in: Tenth International AAAI Conference on Web and Social
  Media, 2016.

\bibitem{li2014towards}
J.~Li, M.~Ott, C.~Cardie, E.~Hovy, Towards a general rule for identifying
  deceptive opinion spam, in: Proceedings of the 52nd Annual Meeting of the
  Association for Computational Linguistics (Volume 1: Long Papers), 2014, pp.
  1566--1576.

\bibitem{you2018attribute}
Z.~You, T.~Qian, B.~Liu, An attribute enhanced domain adaptive model for
  cold-start spam review detection, in: Proceedings of the 27th International
  Conference on Computational Linguistics, 2018, pp. 1884--1895.

\bibitem{kumar2018rev2}
S.~Kumar, B.~Hooi, D.~Makhija, M.~Kumar, C.~Faloutsos, V.~Subrahmanian, Rev2:
  Fraudulent user prediction in rating platforms, in: Proceedings of the
  Eleventh ACM International Conference on Web Search and Data Mining, 2018,
  pp. 333--341.

\bibitem{shehnepoor2017netspam}
S.~Shehnepoor, M.~Salehi, R.~Farahbakhsh, N.~Crespi, Netspam: A network-based
  spam detection framework for reviews in online social media, IEEE
  Transactions on Information Forensics and Security 12~(7) (2017) 1585--1595.

\bibitem{liu2019opinion}
Y.~Liu, B.~Pang, X.~Wang, Opinion spam detection by incorporating multimodal
  embedded representation into a probabilistic review graph, Neurocomputing 366
  (2019) 276--283.

\bibitem{harris2012detecting}
C.~G. Harris, Detecting deceptive opinion spam using human computation, in:
  Workshops at the Twenty-Sixth AAAI Conference on Artificial Intelligence,
  2012.

\bibitem{feng2012syntactic}
S.~Feng, R.~Banerjee, Y.~Choi, Syntactic stylometry for deception detection,
  in: Proceedings of the 50th Annual Meeting of the Association for
  Computational Linguistics (Volume 2: Short Papers), 2012, pp. 171--175.

\bibitem{li2014spotting}
H.~Li, B.~Liu, A.~Mukherjee, J.~Shao, Spotting fake reviews using
  positive-unlabeled learning, Computaci{\'o}n y Sistemas 18~(3) (2014)
  467--475.

\bibitem{kim2015deep}
S.~Kim, H.~Chang, S.~Lee, M.~Yu, J.~Kang, Deep semantic frame-based deceptive
  opinion spam analysis, in: Proceedings of the 24th ACM International on
  Conference on Information and Knowledge Management, 2015, pp. 1131--1140.

\bibitem{mukherjee2012spotting}
A.~Mukherjee, B.~Liu, N.~Glance, Spotting fake reviewer groups in consumer
  reviews, in: Proceedings of the 21st international conference on World Wide
  Web, 2012, pp. 191--200.

\bibitem{li2019unsupervised}
Q.~Li, Q.~Wu, C.~Zhu, J.~Zhang, W.~Zhao, Unsupervised user behavior
  representation for fraud review detection with cold-start problem, in:
  Pacific-Asia Conference on Knowledge Discovery and Data Mining, Springer,
  2019, pp. 222--236.

\bibitem{rayana2015collective}
S.~Rayana, L.~Akoglu, Collective opinion spam detection: Bridging review
  networks and metadata, in: Proceedings of the 21th acm sigkdd international
  conference on knowledge discovery and data mining, 2015, pp. 985--994.

\bibitem{rayana2016collective}
S.~Rayana, L.~Akoglu, Collective opinion spam detection using active inference,
  in: Proceedings of the 2016 SIAM International Conference on Data Mining,
  SIAM, 2016, pp. 630--638.

\bibitem{liu2017holoscope}
S.~Liu, B.~Hooi, C.~Faloutsos, Holoscope: Topology-and-spike aware fraud
  detection, in: Proceedings of the 2017 ACM on Conference on Information and
  Knowledge Management, 2017, pp. 1539--1548.

\bibitem{xu2017online}
C.~Xu, J.~Zhang, Z.~Sun, Online reputation fraud campaign detection in user
  ratings., in: IJCAI, 2017, pp. 3873--3879.

\bibitem{kaghazgaran2018combating}
P.~Kaghazgaran, J.~Caverlee, A.~Squicciarini, Combating crowdsourced review
  manipulators: A neighborhood-based approach, in: Proceedings of the Eleventh
  ACM International Conference on Web Search and Data Mining, 2018, pp.
  306--314.

\bibitem{Wang2020}
Z.~Wang, R.~Hu, Q.~Chen, P.~Gao, X.~Xu, Collueagle: collusive review spammer
  detection using markov random fields, Data Mining and Knowledge Discovery
  34~(6) (2020) 1621--1641.

\bibitem{xu2015combating}
C.~Xu, J.~Zhang, Combating product review spam campaigns via multiple
  heterogeneous pairwise features, in: Proceedings of the 2015 SIAM
  International Conference on Data Mining, SIAM, 2015, pp. 172--180.

\bibitem{xu2013uncovering}
C.~Xu, J.~Zhang, K.~Chang, C.~Long, Uncovering collusive spammers in chinese
  review websites, in: Proceedings of the 22nd ACM international conference on
  Information \& Knowledge Management, 2013, pp. 979--988.

\bibitem{leskovec2010predicting}
J.~Leskovec, D.~Huttenlocher, J.~Kleinberg, Predicting positive and negative
  links in online social networks, in: Proceedings of the 19th international
  conference on World wide web, 2010, pp. 641--650.

\bibitem{hahnloser2001permitted}
R.~H. Hahnloser, H.~S. Seung, Permitted and forbidden sets in symmetric
  threshold-linear networks, in: Advances in neural information processing
  systems, 2001, pp. 217--223.

\bibitem{hahnloser2000digital}
R.~H. Hahnloser, R.~Sarpeshkar, M.~A. Mahowald, R.~J. Douglas, H.~S. Seung,
  Digital selection and analogue amplification coexist in a cortex-inspired
  silicon circuit, Nature 405~(6789) (2000) 947--951.

\bibitem{perozzi2014deepwalk}
B.~Perozzi, R.~Al-Rfou, S.~Skiena, Deepwalk: Online learning of social
  representations, in: Proceedings of the 20th ACM SIGKDD international
  conference on Knowledge discovery and data mining, 2014, pp. 701--710.

\bibitem{mikolov2013efficient}
T.~Mikolov, K.~Chen, G.~Corrado, J.~Dean, Efficient estimation of word
  representations in vector space, arXiv preprint arXiv:1301.3781.

\bibitem{zheng2018fraudne}
M.~Zheng, C.~Zhou, J.~Wu, S.~Pan, J.~Shi, L.~Guo, Fraudne: a joint embedding
  approach for fraud detection, in: 2018 International Joint Conference on
  Neural Networks (IJCNN), IEEE, 2018, pp. 1--8.

\end{thebibliography}

\end{document}